\documentclass[reprint, superscriptaddress]{revtex4-1}
\usepackage{amsthm,amsmath,amssymb}
\usepackage{newtxtext,newtxmath}
\usepackage{booktabs} 
\usepackage[colorlinks,linkcolor=blue,anchorcolor=blue,citecolor=blue,urlcolor=blue]{hyperref}
\usepackage{amsmath,amssymb,amsfonts}%
\usepackage{amsthm}%
\usepackage{mathrsfs}%
\usepackage{xcolor}%
\usepackage{textcomp}%
\usepackage{upgreek}
\usepackage {verbatim}
\usepackage{calc}
\usepackage[export]{adjustbox}
\usepackage{enumitem}
\usepackage{booktabs,array}
\usepackage[T1]{fontenc}
\usepackage{algorithm}%
\usepackage{algorithmicx} %
\usepackage{algpseudocode}%
\usepackage{listings}%
\usepackage{subfigure}
\usepackage{color,bm}
\usepackage{float}
\usepackage[normalem]{ulem}
\usepackage{indentfirst}
\usepackage{diagbox}
\usepackage{graphicx}%
\usepackage{caption}
\newcommand{\rom}[1]{\uppercase\expandafter{\romannumeral#1}}

\allowdisplaybreaks 

\begin{document}

\title{Evidence of scaling advantage on an NP-Complete problem with enhanced quantum solvers}
\affiliation{Beijing Academy of Quantum Information Sciences, Beijing 100193, China}
\affiliation{State Key Laboratory of Low-Dimensional Quantum Physics and Department of Physics, Tsinghua University, Beijing 100084, China}
\author{Quanfeng Lu}
\thanks{These authors contributed equally to this work.}
\affiliation{State Key Laboratory of Low-Dimensional Quantum Physics and Department of Physics, Tsinghua University, Beijing 100084, China}
\affiliation{Beijing Academy of Quantum Information Sciences, Beijing 100193, China}
\author{Shijie Wei}
\email{weisj@baqis.ac.cn}
\thanks{These authors contributed equally to this work.}
\affiliation{Beijing Academy of Quantum Information Sciences, Beijing 100193, China}
\author{Keren Li}
\affiliation{College of Physics and Optoelectronic Engineering, Shenzhen University, Shenzhen 518060, China}
\affiliation{Quantum Science Center of Guangdong-Hong Kong-Macao Greater Bay Area (Guangdong), Shenzhen 518045, China}
\author{Pan Gao}
\affiliation{Beijing Academy of Quantum Information Sciences, Beijing 100193, China}

\author{Bao Yan}
\affiliation{State Key Laboratory of Mathematical Engineering and Advanced Computing, Zhengzhou 450001, China}

\author{Muxi Zheng}
\affiliation{State Key Laboratory of Low-Dimensional Quantum Physics and Department of Physics, Tsinghua University, Beijing 100084, China}

\author{Haoran Zhang}
\affiliation{School of Electrical and Electronic Engineering, Nanyang Technological University, Singapore, Singapore}

\author{Jinfeng Zeng}
\email{zengjf@baqis.ac.cn}
\affiliation{Beijing Academy of Quantum Information Sciences, Beijing 100193, China}

\author{Gui-Lu Long}
\email{gllong@tsinghua.edu.cn}
\affiliation{Beijing Academy of Quantum Information Sciences, Beijing 100193, China}
\affiliation{State Key Laboratory of Low-Dimensional Quantum Physics and Department of Physics, Tsinghua University, Beijing 100084, China}
\affiliation{Frontier Science Center for Quantum Information, Beijing 100084, China}
\affiliation{Beijing National Research Center for Information Science and Technology, Beijing 100084, China}

\begin{abstract}
 Achieving quantum advantage remains a key milestone in the noisy intermediate-scale quantum era. Without rigorous complexity proofs, scaling advantage—where quantum resource requirements grow more slowly than their classical counterparts—serves as the primary indicator.  However, direct applications of quantum optimization algorithms to classically intractable problems have yet to demonstrate this advantage. 
To address this challenge, we develop enhanced quantum solvers for the NP-complete one-in-three Boolean satisfiability problem. We propose a restricting space reduction algorithm (RSRA)  that
achieves optimal search space dimensionality, thereby reducing both qubits  and time complexity for various quantum solvers. Extensive numerical investigations on problem instances with up to 65 variables demonstrate that our enhanced quantum approximate optimization algorithm (QAOA)  and quantum adiabatic algorithm (QAA)-based solvers outperform state-of-the-art classical solvers, with the QAA-based solver providing a lower bound for our method while exhibiting scaling advantage.
Furthermore, we experimentally implement our enhanced solvers on a superconducting quantum processor with  13 qubits, confirming the predicted performance improvements. Collectively, our results provide empirical  evidence of quantum speedup for an NP-complete problem.
\end{abstract}
\maketitle
\section{Introduction}\label{Sec1}
Quantum advantage remains the central objective in quantum computing.
Quantum optimization algorithms, including the quantum approximate optimization algorithm (QAOA)~\cite{farhi2014quantum} and the variational quantum eigensolver (VQE)~\cite{peruzzo2014variational}, are promising candidates for achieving quantum advantage in the noisy intermediate-scale quantum (NISQ) era~\cite{Preskill2018}, which is characterized  by limited qubit counts~\cite{Google2019Supremacy,Wu2021Strong,Kim2023Evidence} and the absence of effective error correction mechanisms \cite{Acharya2023Suppressing,Yu2023Nature,Sivak2023Nature,Lukin2024Nature}.  QAOA balances optimization accuracy with gate usage to mitigate noise, while VQE reduces quantum resource demands through classical optimization. Since these algorithms rely on classical optimization, their time complexities lack rigorous characterization. Demonstrating quantum advantage therefore requires comparing the scalings of their time complexities with those of the best  classical algorithms. To date, systematic studies in this area remain limited. Notably, two significant studies have investigated the scaling advantages of QAOA for the Boolean satisfiability (SAT) problem~\cite{boulebnane2024solving} and the low autocorrelation binary sequences (LABS) problem~\cite{shaydulin2024evidence}.\par 
\begin{table*}\label{scaling factor}
\centering
\sffamily 
    \caption{\textbf{Comparison of scalings in time complexities for different one-in-three SAT solvers across various $\frac{m}{n}$ ratios.}  All QAOA-based solvers utilize 25 layers, while all QAA-based solvers utilize 100 layers. Five types of classical solvers are used to calculate the optimal scaling of classical methods. Scalings lower than those of classical solvers are indicated in boldface. We observe that the enhanced QAOA- and QAA-based solvers  exhibit better scalings than their classical counterparts and quantum methods without classical reduction across the critical region.  $|G|$ denotes the size of the set $G$  which  contains at least two variables in each clause, which is approximately $0.5n$ in the critical region.}	\label{tabscale}
    \renewcommand{\arraystretch}{2.45}
    \setlength{\tabcolsep}{5pt}
\resizebox{0.65\textwidth}{!}{
	\begin{tabular}{>{\huge}c>{\huge}c>{\huge}c>{\huge}c>{\huge}c>{\huge}c>{\huge}c>{\huge}c>{\huge}c>{\huge}c>{\huge}c}
		\toprule[2pt]
        \huge\textbf{$\mathbf{\frac{m}{n}}$ ratio}&\multicolumn{4}{c}{\huge\textbf{Enhanced with the RSRA}}&\vspace{1pt}&\multicolumn{3}{c}{\huge\textbf{Original}}&\vspace{0.5pt}&\huge\textbf{Classical}\\
        
        \cmidrule[1pt]{2-5}\cmidrule[1pt]{7-9}
        &\multicolumn{4}{c}{\huge$\mathbf{|G|(\approx 0.5n)}$ \textbf{qubits}}&\vspace{0.5pt}&\multicolumn{3}{c}{\huge$\mathbf{n}$ \textbf{qubits}}\\
        \cmidrule[1pt]{2-5}\cmidrule[1pt]{7-9}
        &\textbf{QAA}&\textbf{QAOA}&\textbf{VQE}&\textbf{Grover}&
        &\textbf{QAA}&\textbf{QAOA}&\textbf{Grover}\\
		\midrule[1.5pt]
        0.626&\textbf{1.008}&\textbf{1.006}&1.020&1.084&&1.039&1.041&1.345&&1.013\\
0.55&\textbf{1.005}&\textbf{1.002}&1.012&1.098&&1.029&1.021&1.327&&1.012\\
0.575&\textbf{1.006}&\textbf{1.004}&1.013&1.093&&1.032&1.027&1.334&&1.012\\
0.6&\textbf{1.006}&\textbf{1.005}&1.019&1.089&&1.037&1.035&1.340&&1.013\\
0.65&\textbf{1.009}&\textbf{1.008}&1.020&1.078&&1.041&1.037&1.350&&1.013\\
0.675&\textbf{1.009}&\textbf{1.009}&1.025&1.073&&1.043&1.031&1.355&&1.013\\
0.7&\textbf{1.010}&\textbf{1.010}&1.030&1.067&&1.046&1.044&1.359&&1.013\\
0.725&\textbf{1.009}&\textbf{1.011}&1.029&1.061&&1.048&1.043&1.363&&1.013\\
0.75&\textbf{1.008}&\textbf{1.011}&1.020&1.055&&1.049&1.045&1.367&&1.013\\
\bottomrule[2pt]
\end{tabular}  }
\end{table*}
The Boolean satisfiability problem (SAT) is  a central topic in computer science, combinatorics, and logic, with broad applications spanning model checking~\cite{biere1999symbolic, bradley2011sat, mcmillan2003interpolation, amla2005analysis}, combinatorial optimization~\cite{zulkoski2017combining, bejar2007regular, banbara2010generating}, automated deduction~\cite{han2010making, armando2005sat}, and cryptography~\cite{soos2009extending, otpuschennikov2016encoding, massacci2000logical, mironov2006applications}. The 3-SAT problem expressed in conjunctive normal form (CNF), as a conjunction of clauses where each clause is a disjunction of three literals, was the first problem proven to be NP-complete by the Cook-Levin theorem \cite{cook1971}. Notably, 3-SAT can also be expressed in one-in-three form, where each clause is satisfied if and only if exactly one literal is true. This problem is mutually reducible to 3-SAT in CNF and admits a simpler Hamiltonian representation, making it well-suited for quantum optimization (see Section S1 in Supplementary Information). To the best of our knowledge, no direct classical solver exists for the one-in-three SAT problem; existing classical methods typically rely on reductions to exact cover or SAT in CNF. These characteristics make the one-in-three SAT problem a compelling candidate for demonstrating quantum advantage over classical approaches.\par
Several quantum algorithms have been proposed for solving the  SAT problem using various quantum ground state solvers, including quantum annealing (QA)~\cite{nusslein2023solving}, the quantum adiabatic algorithm (QAA)~\cite{albash2018adiabatic,hogg2003adiabatic, wang2016ultrafast, farhi2001quantum,young2008size,young2010first}, and the quantum approximate optimization algorithm (QAOA)~\cite{farhi2014quantum,golden2023quantum, mandl2024amplitude, bengtsson2020improved}. A recent numerical study~\cite{boulebnane2024solving}  demonstrated that the time complexity of  QAOA with fixed parameters and constant depth, obtained from data training, for random 8-SAT problems with $n\leq20$ variables scales as $1.23^n$. 
However, these quantum algorithms are direct implementations of quantum optimization techniques applied to the SAT problem without optimizing the dimensionality of the search space, which can lead to large scalings or difficulties in parameter training~\cite{boulebnane2024solving}, hindering their ability to demonstrate  quantum advantage over classical methods.\par
In this paper, we introduce enhanced quantum one-in-three SAT solvers that exhibit quantum scaling advantage. We propose the restricting space reduction algorithm (RSRA), which reduces the number of candidate solutions from
$2^n$ to approximately $2^{n-m}$, where $n$ denotes the number of variables and $m$ represents the number of clauses. By integrating the RSRA with several quantum optimization algorithms, we develop enhanced VQE-, QAOA-, and QAA-based solvers with reduced time  complexities. These enhanced quantum solvers exhibit the following key features:
\begin{itemize}
    \item[(1)] \textbf{Optimal search space reduction:} The RSRA reduces the search space dimensionality from $2^n$ to approximately $2^{n-m}$.
    \item[(2)] \textbf{Problem heuristic  ansatz:} We utilize the RSRA to construct problem heuristic  ansatz that ensure quantum states remain within the subspace spanned by solutions to the loosened problem while simultaneously simplifying the problem Hamiltonian. Compared to the original ansatz, the new ansatz demonstrate improvements in qubit usage and search space dimensionality. Additionally, we prove that the enhanced VQE-based solver can  avoid barren plateaus.
    \item[(3)] \textbf{Scalability in simulation and experiment:} The RSRA  enables large-scale numerical simulations and more feasible experimental demonstrations of SAT instances. Specifically, our VQE-based solver can handle instances with up to $n=150$ variables, while our QAOA- and QAA-based solvers scale to problems with up to $n=100$ variables. Experimentally, we demonstrate the feasibility of our approach on a superconducting quantum processor by using 13 qubits to solve instances with 20 variables, achieving meaningful results.
    \item[(4)] \textbf{Quantum scaling advantage:} The enhanced solvers exhibit clear scaling advantages. The QAOA-based solver achieves the best performance, while the QAA-based solver provides a  lower bound. For problem sizes up to $n=65$ variables at the critical ratio $\frac{m}{n}=0.626$, the enhanced QAOA-based solver with 25 layers and the QAA-based solver with 100 layers scale as $1.006^n$ and $1.008^n$, respectively, both outperforming the $1.013^n$ scaling of state-of-the-art classical solvers.
\end{itemize}\par
In summary, these advances substantially reduce the barriers to demonstrating quantum scaling advantages for an NP-complete problem in the NISQ era.\par
	\begin{figure*}
		\centering
		\includegraphics[width=0.805\linewidth]{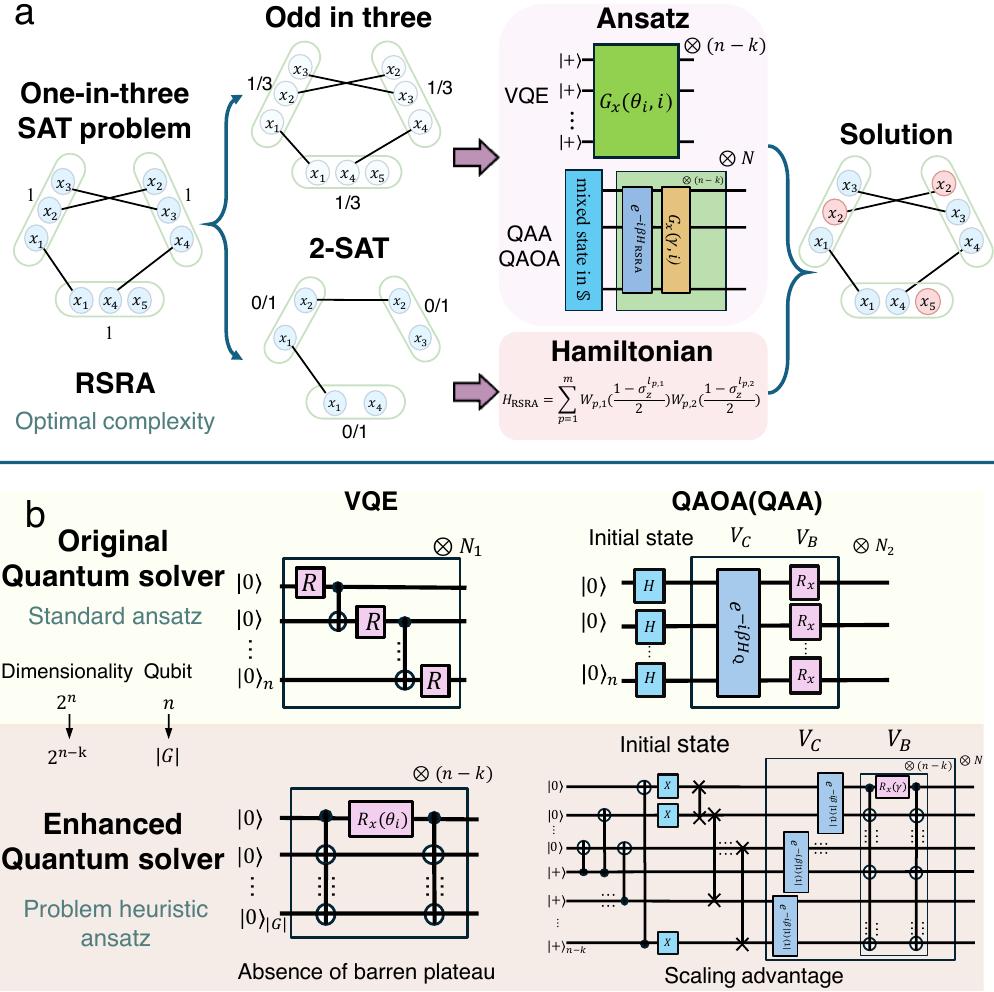}
		\caption{ \textbf{Schematic illustration of the RSRA and the enhanced quantum solvers.} \textbf{a} An overview of the RSRA and its applications to quantum solvers for the one-in-three SAT problem.  The RSRA transforms the original problem into a 2-SAT problem within the solution space of the loosened problem, enabling the construction of an efficient Hamiltonian and various ansatz that preserve the reduced solution space. This leads to enhanced solvers based on VQE, QAA, and QAOA, potentially achieving optimal time complexity. \textbf{b} Comparison of different ansatz designs for the original and enhanced quantum solvers. The standard ansatz used in the original quantum solvers are illustrated in the upper half, while the problem  heuristic ansatz adopted in the enhanced quantum solvers are depicted in the lower half. For all enhanced solvers, the search space dimensionality is reduced from $2^n$ to $2^{n-k} = \mathcal{O}(2^{n-m})$, and the number of required qubits is reduced from $n$ to $|G|$. Moreover, the enhanced VQE-based solver avoids barren plateaus, while the enhanced QAOA- and QAA-based solvers exhibit a clear scaling advantage.
.} 
		\label{fig:workflow}
	\end{figure*}

\section{Results}
\subsection{One-in-three SAT problem}
We first define the one-in-three SAT problem and review its existing classical and quantum solvers. In this problem, each clause requires exactly one of its three literals to be true. The objective is to determine an assignment that satisfies all clauses, or to determine that no such assignment exists.  Let $m$ and $n$ denote the number of clauses and  variables, respectively.  In random cases~\cite{farhi2001quantum},  clauses are generated by selecting  literals randomly from the set $\{1,2,\cdots, n\}$, so the actual number of  variables involved in the clauses may be smaller than $n$. We use $C_{p, i}$ and $x_{p, i}$ to denote the $i$-th literal and variable in the $p$-th clause, respectively. The corresponding label is denoted as $l_{p, i}$, where $x_{l_{p, i}} = x_{p, i}$.  The function $W_{p, i}(x)$ is defined as $x$ or $1-x$, depending on whether the literal is in positive or negative form, such that $W_{p, i}(x_{p, i})= C_{p, i}$. In positive one-in-three SAT problems where all literals are in positive form, a critical point of clause-to-variable ratio $\frac{m}{n}$ exists at $0.626$ \cite{raymond2007phase}, far from which instances are relatively easy to solve. Therefore, this study focuses on positive one-in-three SAT problems with clause-to-variable ratios ranging from $0.55$ to $0.75$.\par
To our knowledge, no  direct classical solver exists for the one-in-three SAT problem. This problem is typically addressed by reducing it to a 3-SAT problem in CNF, which can be solved using standard SAT solvers, or to an exact cover problem, which can be efficiently solved using the Dancing Links X (DLX) algorithm. 
In existing quantum  SAT solvers,  variables are encoded  onto  qubits  as $
x_{i}=\frac{1-\sigma^{i}_z}{2},x_{p,i}=\frac{1-\sigma^{l_{p,i}}_z}{2}$.
The one-in-three SAT problem can then be addressed by minimizing the following Hamiltonian:
\begin{align}
	H_{\rm{Q}}=&\sum\limits_{p=1}^{m}(C_{p,1}+C_{p,2}+C_{p,3}-1)^2\notag\\
	=&\sum\limits_{p=1}^{m}(W_{p,1}(x_{p,1})+W_{p,2}(x_{p,2})+W_{p,3}(x_{p,3})-1)^2\notag\\
	=&\sum\limits_{p=1}^{m}(W_{p,1}(\frac{1-\sigma_z^{l_{p,1}}}{2})+W_{p,2}(\frac{1-\sigma_z^{l_{p,2}}}{2})\notag\\
	&+W_{p,3}(\frac{1-\sigma_z^{l_{p,3}}}{2})-1)^2,
\end{align} 
using various quantum ground state solvers, each offering a distinct quantum solver. Solvers based on the QAA typically reach the ground state with high probability. In contrast, hybrid solvers based on the VQE and the QAOA  are more practical for NISQ devices, though they require classical optimization and may not always converge to the ground state. 

\subsection{Enhanced quantum  SAT solvers employing classical reduction}\label{novel}
We propose a classical reduction algorithm: the restricting space reduction algorithm (RSRA). It transforms the original one-in-three SAT problem into a 2-SAT problem within the solution space of the `loosened problem', obtained by relaxing clauses from `one-in-three' to `odd-in-three'. Each independent clause under modular 2 reduces the search space dimensionality by half, leading to an optimal search space dimensionality. We demonstrate that the time complexity  of the RSRA is  polynomial in the problem size (see Section S2 in the Supplementary Information).
As shown in Fig.~\ref{fig:combine1}a and Fig.~\ref{fig:combine1}b, 
the average values of $k$  exhibit linear relationships with problem size $n$ when $\frac{m}{n}$  is fixed and their gradients are approximately $\frac{m}{n}$ in the range between 0.55 and 0.75. Consequently, the average values of $k$ are close to  $m$ within the critical region.
	\begin{figure*}
		\centering
		\includegraphics[width=1\linewidth]{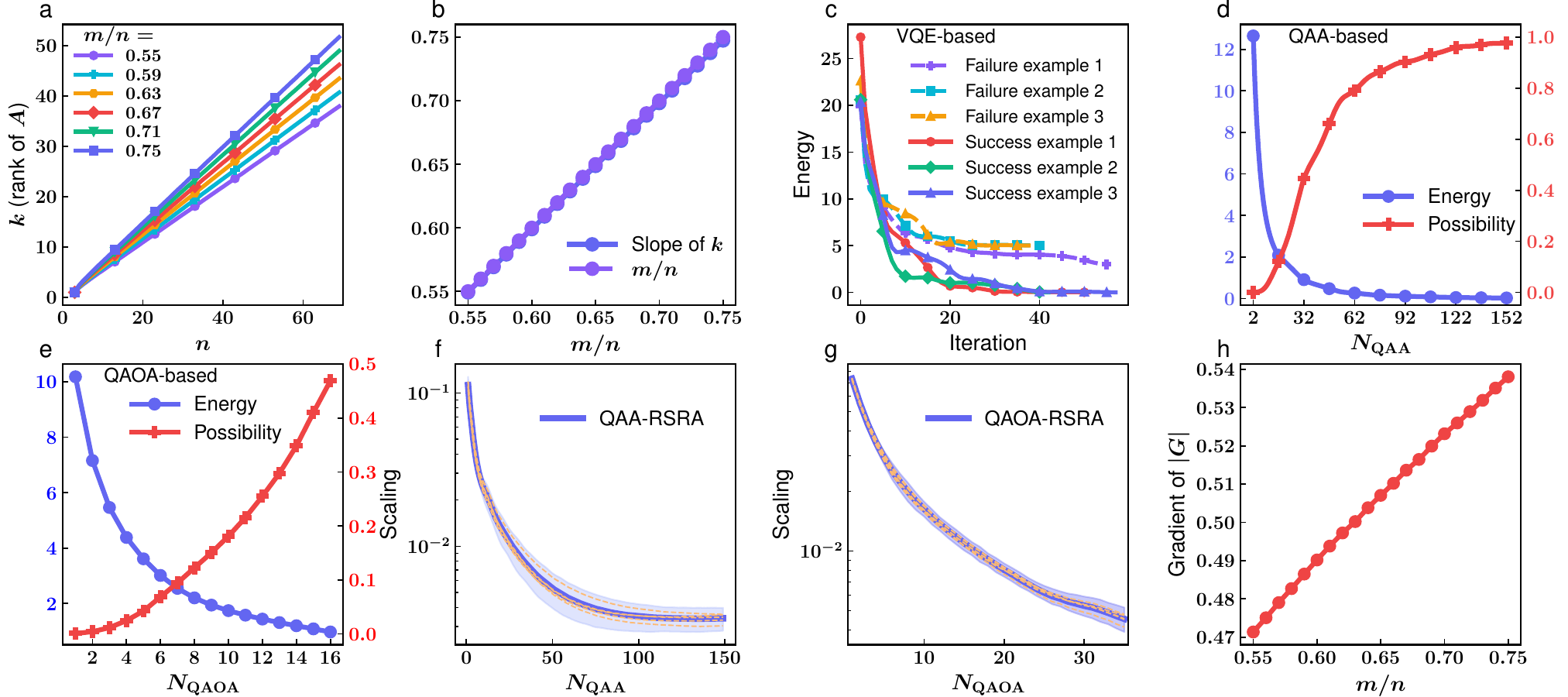}
		\caption{\textbf{The numerical simulation results.
       } 
         \textbf{a} Average values of $k$ under fixed $\frac{m}{n}$ ratios across varying problem sizes $n$,  revealing a linear correlation between the average  values of $k$ and $n$ across different $\frac{m}{n}$ ratios.  \textbf{b} Slopes of average $k$ with respect to $n$ across  different $\frac{m}{n}$ ratios, compared to the line  $y=\frac{m}{n}$. Notably, in the critical region, the slopes of $k$  are closely aligned with $\frac{m}{n}$, suggesting that $k\approx m$.  \textbf{c} Three examples of successful runs and three examples of failed attempts from the enhanced VQE-based  solver  on a  SAT problem with 150 variables and 94 clauses. \textbf{d} Energy and success possibility curves of the enhanced QAA-based solver with respect to the number of layers $N_\mathrm{QAA}$, when applied to a SAT problem with 100 variables and 63 clauses. \textbf{e} Energy and success possibility curves of the enhanced QAOA-based solver with respect to the number of layers $N_\mathrm{QAOA}$, when applied to the same SAT problem with 100 variables and 63 clauses. \textbf{f} and \textbf{g} Scalings of QAA- and QAOA-based solvers for problem sizes up to 40 qubits, as a function of the number of layers. Solid lines represent the average scalings, while dashed lines indicate results from five individual runs. The shaded regions represent 95\% confidence intervals calculated from these runs. \textbf{h} Slopes of average $|G|$ in the critical region, indicating that $|G|$ is approximately $0.5n$ within this region.}
	\label{fig:combine1}
\end{figure*}
 By applying the RSRA, we obtain a quantum solver with a time complexity of $\mathcal{O}(\sqrt{2^{n-m}})$ by preparing a mixed state in the solution space of the loosened problem (see Subsection S3 A in the Supplementary Information) and employing Grover's search algorithm. 
To meet the demands of the NISQ era and potentially reduce time complexity, we also propose enhanced solvers based on  VQE, QAA, and QAOA.
The RSRA facilitates the construction of ansatz that constrain the quantum state to the solution space of the loosened problem. The one-in-three SAT problem can be solved by employing these ansatz and identifying  the ground state of the Hamiltonian corresponding to the 2-SAT problem using VQE, QAA, or QAOA. The relevant Hamiltonian can be expressed as: 
\begin{eqnarray}\label{eq:H20}
	H_{\mathrm{RSRA}}=\sum\limits_{p=1}^{m} W_{p,1}(\frac{1-\sigma_z^{l_{p,1}}}{2})
	W_{p,2}(\frac{1-\sigma_z^{l_{p,2}}}{2}).
\end{eqnarray}
Due to their optimal search space dimensionality, these enhanced solvers have the potential to achieve  optimal time complexity.

\subsection{Numerical simulation of  SAT problem}
In this subsection, we present the classical simulation results for solving randomly generated positive one-in-three SAT problems using the enhanced quantum SAT solvers, highlighting their considerable success possibilities. The RSRA enables us to simulate instances  with 150 variables, which is challenging for existing quantum SAT solvers. The results are illustrated in Fig.~\ref{fig:combine1}c-e. Although the enhanced solvers are applicable to general one-in-three SAT problems, we restrict all clauses to contain only positive literals to ensure that the corresponding `loosened problem' has at least one solution.\par
First, we randomly generate a one-in-three SAT problem with 150 variables and 94 clauses, in which 131 variables actually occur.
We solve this instance using the enhanced VQE-based solver with random initialization and  the Nesterov-accelerated Adaptive Moment Estimation (Nadam) optimizer. The expectation value of the Hamiltonian is calculated by multiplying $4\times4$ matrices (see Subsection S3 D in Supplementary Information). We terminate the simulation when the norm of the gradient vector falls below 0.1, and determine success if the final energy is less than 0.5. In our experiments, 166 out of 2000 attempts succeed, yielding a success probability of $8.3\%$, which is notable considering the problem size. We plot the energy curves for three successful examples and three failed attempts in Fig.~\ref{fig:combine1}c.\par
Next, we randomly generate a SAT instance with 100  variables and 63 clauses, in which 86 variables actually occur. Fig. \ref{fig:combine1}d shows that, as the number of layers $N_{\mathrm{QAA}}$ increases from 2 to 152, the energy drops from above 12 to below 0.1, while the probability of finding a satisfying assignment rises from nearly zero to more than 0.9. We apply our enhanced QAOA-based solver to the same instance and optimize the circuit parameters with the Broyden–Fletcher–Goldfarb–Shanno (BFGS) optimizer. As $N_{\mathrm{QAOA}}$ increases from 1 to 16, the minimum energy  decreases from above 10 to below 1, and the corresponding success probability increases from below 0.1 to above 0.4 as shown in Fig.~\ref{fig:combine1}e.
\subsection{Scalings  of  time complexities }\label{scaling}
In this subsection, we evaluate the performance of our enhanced solvers on random instances, following the standard convention in SAT problem research concerning scaling  \cite{boulebnane2024solving,young2008size,young2010first}.
Scaling is defined as the  factor $c^{n}$ in the average time complexity for randomly generated cases, serving as a quantitative measure of solver efficiency. Time complexity is assessed by estimating the exponentially small average success probabilities of quantum solvers and the exponentially large average number of conflicts encountered by classical solvers. By performing linear regressions of the logarithms of both quantities as functions of problem size $n$, we  determine the values of $c$. \par
For QAA-based solvers, as well as for initializing QAOA-based solvers, parameter sets are chosen according to the principles of the quantum adiabatic algorithm.
 In adiabatic quantum computing, the equivalent Hamiltonian is given by  $H(f(s))=(1-f(s))H_{0}+f(s)H_{1}$, where the function $f$ is referred to as the scheduling function. It is a strictly increasing mapping on the interval $[0,1]$ with $f(0)=0$, $f(1)=1$. In this study, $f$  is chosen as :
\begin{equation}\label{eq:schedule}
f(s)=c_e^{-1}\int_0^s\exp\left(-5{s^{\prime}(1-s^{\prime})}\right)\mathrm{d}s^{\prime},
\end{equation}
where $c_{e}=\int_{0}^{1}\exp{(-5s^{\prime}(1-s^{\prime}))}\mathrm{d}s^{\prime}$ is a normalization constant ensuring that $f(1)=1$. This scheduling function ensures  that the equivalent Hamiltonian evolves rapidly at both ends but slows down in the middle, where exponentially small gaps may arise. The parameter sets are constructed as discrete approximations of $f$, with integrals replaced by summations. The duration of each discrete step is set to 0.25 for QAA-based solvers, and 0.5 for the initial parameters of QAOA-based solvers.\par
To evaluate the required number of layers, we first analyze the average success possibility of the enhanced QAA- and QAOA-based solvers with increasing  layer counts, averaged over 2000 cases. Specifically, we  calculate the scalings for QAA-based solvers up to $150$  layers and for QAOA-based solvers up to $35$  layers at the critical ratio $\frac{m}{n}=0.626$. The data are collected for problem sizes up to 40, which can be efficiently simulated classically. As shown in Fig.~\ref{fig:combine1}f and Fig.~\ref{fig:combine1}g, the scalings for the enhanced QAA- and QAOA-based solvers can be  reduced to $1.003^n$ and $1.005^n$, respectively, indicating their potential to outperform classical counterparts. The performance of the enhanced QAA-based solver does not improve significantly for layer numbers exceeding 100; therefore, we select 100 layers for subsequent simulations of the QAA-based solver. \par
 For QAA-based solvers, the performance is  evaluated over 10000 cases. For QAOA-based solvers, the same parameter sets are initially selected, optimized using 500 training cases, and evaluated over 2000 test cases. For VQE-based solvers, the average success possibility is obtained from 1000 test cases, with random initialization and employing the Nadam optimizer. In classical methods, one-in-three SAT problems are typically reduced to either SAT problems in CNF or exact cover problems. The former can be efficiently solved using state-of-the-art classical SAT solvers, such as MiniSAT \cite{sorensson2005minisat,sorensson2010minisat}, Glucose \cite{audemard:hal-03299473}, Lingeling \cite{Biere-SAT-Competition-2017-solvers,Biere-SAT-Competition-2016-solvers} and Cadical \cite{queue2019cadical}, while the latter is addressed with the Dancing Links X algorithm (DLX) \cite{knuth2000analogue}. For each problem size, 5000 instances are generated to estimate the average conflict time for all classical methods, simulated by Python  package~\textsc{pysat}. For Grover-based solvers, data are obtained by calculating the square roots of the average search space dimensionalities. Data for different solvers are collected for different maximum problem sizes and are appropriately truncated to mitigate polynomial factors. Data for QAA-based and QAOA-based solvers are collected for $n$ up to 65 and truncated at 45. Data for QAA- and QAOA-based solvers without classical reduction are only collected for $n$ between 5 and 12. Data for classical and VQE-based solvers are collected for $n$ up to 120 and truncated at 70. \par
The results for all solvers at  various $\frac{m}{n}$ ratios are summarized in Table~\ref{tabscale}, which consolidates the data presented in Fig. S2-S10 in the Supplementary Information. Notably, the QAA-based solver with 100 layers and the QAOA-based solver with 25 layers, both utilizing the RSRA, consistently outperform classical solvers.  In contrast, solvers without classical reduction do not scale better than their classical counterparts. These findings demonstrate the scaling advantages of the enhanced solvers over classical counterparts and quantum methods without classical reduction, highlighting their potential to achieve quantum advantage in solving one-in-three SAT problems.\par
\subsection{Experiment on superconducting quantum processors}\label{experimental results}
To highlight the advantages of our enhanced solvers for near-term quantum hardware, we experimentally implement our enhanced solvers on a positive one-in-three SAT instance with eight variables and five clauses.
The computations are performed on a superconducting quantum processor comprising  13 transmon qubits, of which 8 qubits are used. Each qubit is coupled to its nearest neighbors via frequency-tunable couplers. The processor's topological structure and compilation method are illustrated in Fig.~\ref{fig:processor}, with further details provided in Table S1 in the Supplementary Information. \par We implemented four schemes to address this SAT problem: a VQE-based solver, an enhanced VQE-based solver, a QAOA-based solver, and an enhanced QAOA-based solver. We maintained the same number of parameters across all VQE-(QAOA-) based solvers, and the gate counts for each circuit are summarized  in Table S2  in the Supplementary Information. 
The detailed form of the SAT problem, the matrix $L$, the QUBO matrix, and the correspondingly constructed ansatz are provided in Fig. S11 in the Supplementary Information. Each run consists of 6 iterations, and all results are averaged over five optimization runs. To compute the gradients of the Hamiltonian with respect to a parameter, we employ the parameter shift technique \cite{li2017hybrid,mitarai2018quantum} for the VQE-based solvers and finite differences with a step size of 0.2 for the QAOA-based solvers.\par
\begin{figure*}
	\centering
	\includegraphics[width=0.95\linewidth]{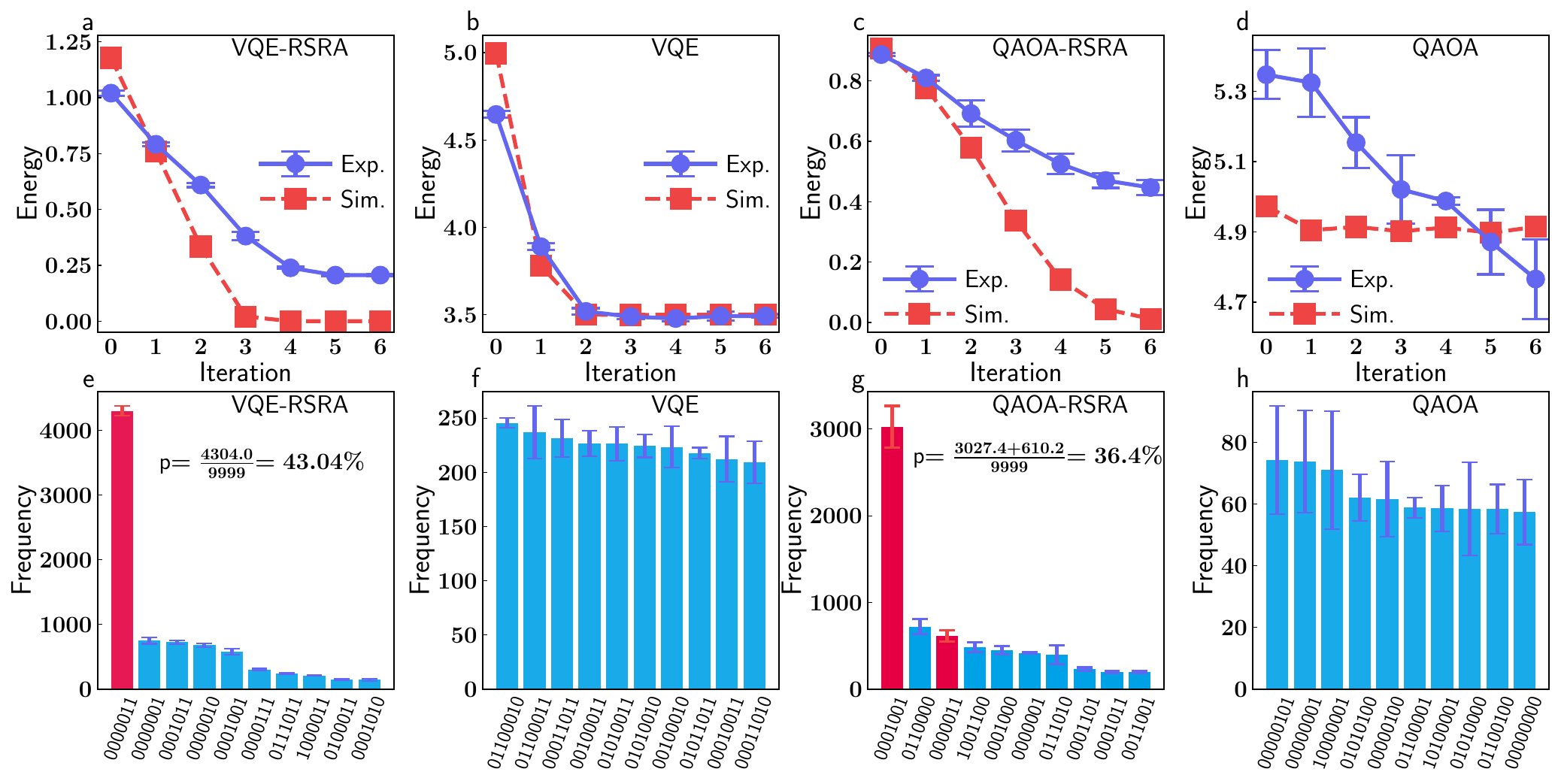}
	\caption{\textbf{Average energy and possibility curves  for VQE- and QAOA-based solvers.} \textbf{a} Energy curve for the enhanced VQE-based solver. \textbf{b} Energy curve for the VQE-based solver without classical reduction. \textbf{c} Energy curve for the enhanced QAOA-based solver. \textbf{d} Energy curve for the QAOA-based solver without classical reduction.\textbf{e} Possibility curve for the enhanced VQE-based solver. \textbf{f} Possibility curve for the VQE-based solver without classical reduction. \textbf{g} Possibility curve for the enhanced QAOA-based solver. \textbf{h} Possibility curve for the QAOA-based solver without classical reduction.
    Curves labeled `Exp.' correspond to the energy curve measured on a quantum processor, whereas curves labeled `Sim.' represent the energy curve obtained through classical simulation. All results are averaged over 5 independent optimization runs to enhance statistical reliability.}\label{fig:exp1}
\end{figure*}

Fig.~\ref{fig:exp1} displays the $\langle H\rangle$ curves and final possibility distributions for all VQE- and QAOA-based solvers. The enhanced VQE- and QAOA-based solvers  achieve final energies of approximately 0.2 and 0.5, respectively. In contrast, the final energies of the VQE- and QAOA-based solvers without classical reduction remain above 3.5. Notably, the final energies of the solvers without classical reduction do not approach 0 even under classical simulation, further demonstrating the effectiveness of our classical reduction method. The success probabilities for the enhanced VQE-based solver and the QAOA-based solver are $43.04\%$ and $36.40\%$, respectively. In contrast, the solvers without classical reduction exhibit nearly zero success probabilities, with none of the solutions appearing among their 10 most frequent outcomes in the distribution.  Detailed data, including changes in expectation of the Hamiltonian and parameter values for all experiments on the quantum processor, are provided  in Tables S3-S10 in the Supplementary Information. These results confirm that the enhanced solvers can achieve significantly higher success possibilities than their counterparts without classical reduction, thereby highlighting their effectiveness.
\begin{figure*}
	\centering
	\includegraphics[width=0.8\linewidth]{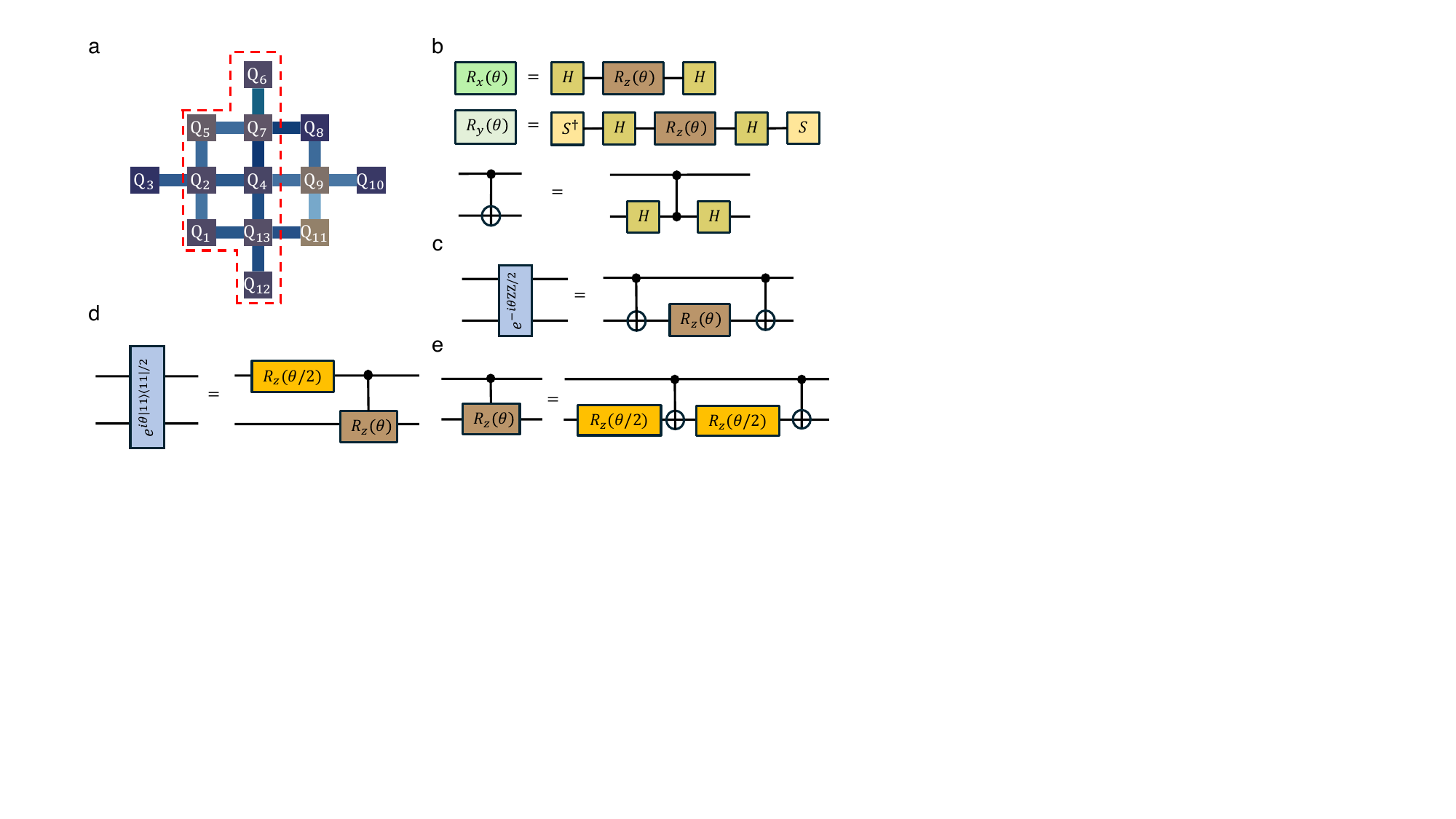}
	\caption{\textbf{Topological structure of  the quantum processor and its compilation method.} \textbf{a} Topological layout of the  quantum processor. The colors of the qubits and the connections between pairs of qubits indicate their single-qubit and two-qubit gate fidelities, respectively. In the first experiment, the employed qubits are indicated by red dashed boxes, whereas the second experiment involves all 13 qubits. \textbf{b} Compilation methods for $R_x$ gates, $R_y$ gates and \textsc{cnot} gates. \textbf{c} Decomposition of $e^{-i\theta ZZ/2}$ into two \textsc{cnot} gates and a $R_z$ gate.  \textbf{d} Decomposition of $e^{i\theta|11\rangle\langle11|/2}$ into an $R_z$ gate and a controlled-$R_z$ gate,  with the latter further decomposed into two \textsc{cnot} gates and two $R_z$ gates as shown in \textbf{e}. }
	\label{fig:processor}
\end{figure*}
\par
\begin{figure*}
	\centering
	\includegraphics[width=0.62\linewidth]{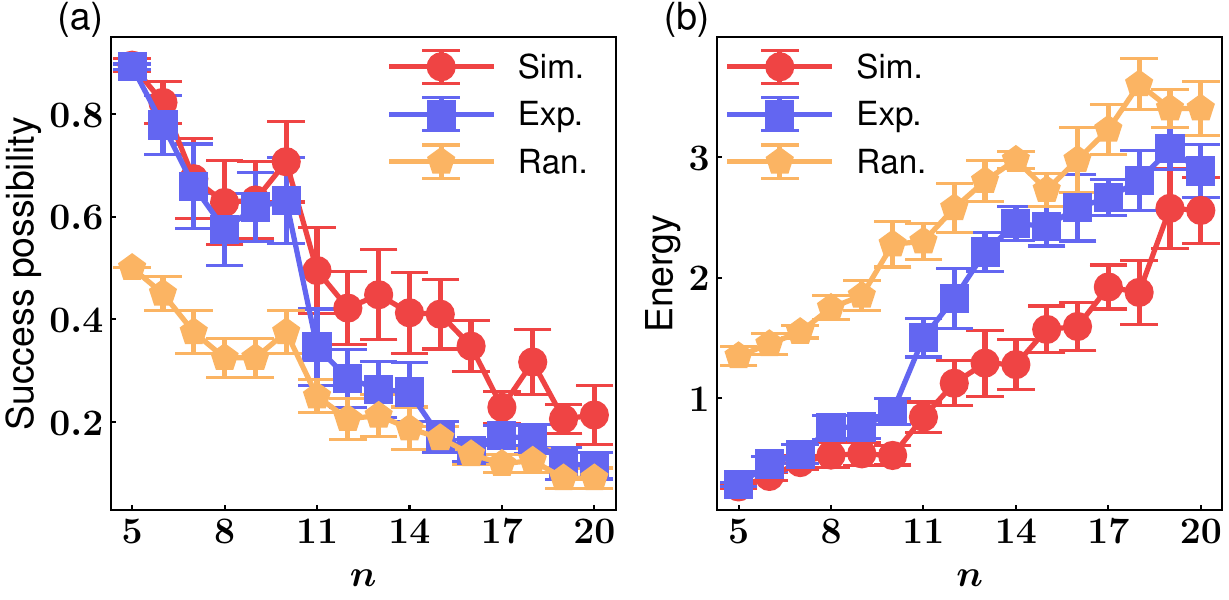}
	\caption{\textbf{Experimental results of the enhanced QAOA-based solver with fixed parameters on a quantum processor involving up to 20 variables.} \textbf{a} Average success possibility for instance of different problem sizes. \textbf{b} Average energy for instances of different problem sizes.  `Sim.' denotes  results obtained from numerical simulations, while `Exp.' refers to experimental results. `Ran.' denotes outcomes produced by random selection from the solution space of the loosened problem. All results are averaged over 10 instances, and error bars represent the standard error of the mean.}
	\label{fig:exp2}
\end{figure*}

 To further validate the feasibility of our approach on larger scale problems, we experimentally implement the enhanced QAOA-based solver with a single layer and fixed parameters.  We generate 10 problem instances for each problem size $n$ ranging from 5 to 20. Fixed parameters for each problem size have been determined through classical simulations conducted during the evaluation of scalings. The average success probability and energy are presented in Fig.~\ref{fig:exp2}a and Fig.~\ref{fig:exp2}b, respectively.
We demonstrate both experimental and simulated results and use randomly selected states from the solution space of the `loosened problem' as a performance baseline. First, our solver  outperforms random guessing in terms of both achieved energy and success probability. Second, random guessing provides a performance guarantee. Notably, for larger problems (e.g. $n=20$), the standard QAOA-based solver fails to produce meaningful results.

\section{Discussion} \label{conclu}
This study introduces  enhanced quantum solvers for one-in-three SAT problems and empirically demonstrates their scaling advantage. We propose the RSRA, which utilizes clause structures to reduce the search space from $2^{n}$ to $\mathcal{O}(2^{n-m})$.  By constructing ansatz that preserve the reduced subspace and minimizing the energy, we develop enhanced VQE-, QAA-, and QAOA-based solvers. Our numerical simulations demonstrate that the enhanced QAA-based solver with $100$ layers and the QAOA-based solver with $25$ layers outperform classical solvers and quantum methods without classical reduction. Additionally, we implement our enhanced solvers on a superconducting quantum processor to demonstrate their advantages over counterparts without classical reduction, as well as their practical feasibility.  These results highlight the potential  of our algorithms to achieve quantum speedup in the NISQ era on an NP-complete problem. \par

The scaling estimates of various solvers presented in this work are derived from extensive calculations for problem sizes up to 120  variables, representing the maximum scale achievable within our current computational capability. At this scale, our simulations indicate that the enhanced quantum solvers demonstrate superior scaling advantages in time complexity compared to classical counterparts. Three observations support our conclusion. First, the coefficients of determination for most solvers are relatively high after truncation, suggesting  that the influence of polynomial factors has been largely mitigated. Second, the scaling of our QAA- and QAOA-based solvers could be further reduced through amplitude amplification \cite{brassard2000quantum} and better schedule functions, while the performance of our VQE-based solver can be improved via more effective classical optimizers or better ansatz design. Third, the parameters for  QAA-based solvers are deterministically chosen according to specific rules rather than through classical optimization, resulting in more reliable scaling estimates.
\par

By employing the RSRA, we transform the one-in-three SAT problem into a 2-SAT problem within the solution space of the `loosened problem'. As demonstrated in \cite{schoning1999probabilistic,bravyi2006efficient},  2-SAT and quantum 2-SAT problems can  be solved in polynomial time. The idea of leveraging problem-specific properties to transform the original problem into a loosened problem for dimensionality reduction can be extended to general SAT problems.
We are exploring more efficient methods for solving 2-SAT problem while maintaining the state within the solution space of the loosened problem. For instance, we may employ the Quantum Monte Carlo method or quantum imaginary time evolution  techniques to implement $e^{-\beta H}$ on the mixed state in the subspace,  thereby finding the solution to the original one-in-three SAT problem.

\section{methods}

\subsection{Restricting  space reduction algorithm}\label{reduction2}
	We observe that existing SAT solvers have not fully exploited the independence of clauses as constraints. To enhance the efficiency of quantum solvers, we propose a classical reduction method—the restricting space reduction algorithm (RSRA). This algorithm transforms the original one-in-three SAT problem into a 2-SAT problem within the solution space of the `loosened problem', obtained by relaxing clauses from `one-in-three' to `odd-in-three'. Each independent clause under modulo 2 reduces the search space dimensionality by half, leading to optimal search space dimensionality. \par
The `loosened problem' is defined by loosening the clause restrictions from `one-in-three' to `odd in three'. If all three literals in a clause are true, the loosened clause is also satisfied.  Accordingly, the $p$-th loosened clause can be expressed as $C_{p, 1}+C_{p, 2}+C_{p, 3}\equiv1$, where $\equiv$ denotes equivalence under modulo 2.
We demonstrate that the solutions to the `loosened problem' can be expressed in the following form (see Section S2 in the Supplementary Information  for more details):
	\begin{equation}\label{eq:form}
		\begin{pmatrix}
			x_1\\
			x_2\\
			\vdots\\
			x_{n}
		\end{pmatrix}
		\equiv
		\begin{pmatrix}
			L_{11}&L_{12}&\cdots&L_{1\:n-k}\\
			L_{21}&L_{22}&\cdots&L_{2\:n-k}\\
			\vdots&\vdots&\ddots&\vdots\\
			L_{n1}&L_{n2}&\cdots&L_{n\:n-k}
		\end{pmatrix}	
		\begin{pmatrix}
			S_1\\
			S_2\\
			\vdots\\
			S_{n-k}
		\end{pmatrix}
		+\begin{pmatrix}
			T_1\\
			T_2\\
			\vdots\\
			T_n
		\end{pmatrix}.
	\end{equation}
In this equation, $L$ is a Boolean matrix and $T$ is a Boolean vector, both  derived from the one-in-three SAT problem, while $S$ is any $(n-k)$-dimensional Boolean vector. Here, $k$  equals the number of independent clauses  under modulo 2. Thus, we can construct a new representation $\mathbb{S}$ of the quantum state within the solution space of the `loosened problem', given by:
	\begin{equation}
		|x_1x_2\cdots x_n\rangle_{\in\;\text{solution\;space}}=|S_1S_2\cdots S_{n-k}\rangle_{\rm{\mathbb{S}}}.
	\end{equation}
	Since $S_1, S_2\cdots, S_{n-k}$ can be freely chosen from the set $\{0,1\}$, the dimensionality of the solution space for the loosened problem is $2^{n-k}$.\par
We impose the following restriction: two literals are selected from each clause and cannot  be assigned 1 simultaneously. A solution to the loosened problem is a solution to the original one-in-three SAT problem if and only if it satisfies this constraint. This constraint can be expressed as a 2-SAT problem, whose $p$-th clause is given by:
	\begin{equation}
	    \neg(C_{p,i}) \lor \neg(C_{p, j}) =1,i\neq j\in(1,2,3).
	\end{equation}
	In this way, the RSRA  reduces the original problem to solving a 2-SAT problem within the solution space  of the `loosened problem',  which has a dimensionality of $2^{n-k}$.
 We demonstrate in Section S2 in the Supplementary Information that the time complexity  of the RSRA is  in polynomial in the problem size.
As shown in Fig.~\ref{fig:combine1}a and Fig.~\ref{fig:combine1}b, 
the average values of $k$  exhibit linear relationships with problem size $n$ when $\frac{m}{n}$  is fixed, and their slopes are approximately equal to $\frac{m}{n}$ in the range between 0.55 and 0.75. Consequently, the average values of $k$ are close to  $m$ within the critical region, indicating that the RSRA can reduce the dimensionality of the search space from $2^n$ to approximately $2^{n-m}$.
\subsection{Applications to  quantum SAT solvers}
  The RSRA immediately yields a quantum solver with  time complexity of $\mathcal{O}(\sqrt{2^{n-m}})$ by preparing a mixed state in the $\mathbb{S}$ representation (see  Subsection S3 C in the Supplementary Information) and applying Grover's search. The RSRA also facilitates the construction of ansatz that ensure the quantum state remains within the solution space of the loosened problem. Consequently, a solution to the one-in-three SAT problem can be obtained by employing these ansatz and identifying the ground state of the Hamiltonian corresponding to the 2-SAT problem using corresponding quantum ground state solvers, including VQE, QAA, and QAOA. These solvers operate within search spaces with optimal dimensionality, potentially achieving optimal time complexity.\par
All enhanced quantum solvers  employ the same encoding as existing quantum SAT solvers and share the same Hamiltonian:
\begin{equation}\label{eq:H2}
	H_{\mathrm{RSRA}}=\sum\limits_{p=1}^{m} W_{p,1}(\frac{1-\sigma_z^{l_{p,1}}}{2})
	W_{p,2}(\frac{1-\sigma_z^{l_{p,2}}}{2}),
\end{equation}
where $p,1$ and $p,2$ can be  freely replaced by $p,3$. This Hamiltonian imposes a penalty term when a clause contains two simultaneously true literals. Given the form of $H_{\mathrm{RSRA}}$, we can select a set $G$ that is as small as possible while containing at least two variables from each clause. Consequently, $H_{\mathrm{RSRA}}$ can  be expressed in terms of the variables in $G$, thereby reducing the number of required qubits to the size of the set $G$, denoted by $|G|$. Since each clause contains at most one variable not belonging to $G$, the values of all variables can be uniquely recovered. With appropriate selection, the average value of $|G|$ can approach $0.5n$ within the critical $\frac{m}{n}$ region between 0.55 and 0.75 as demonstrated in Fig. \ref{fig:combine1}h.\par
As a preliminary step, we  construct several fundamental gates in the $\mathbb{S}$ representation for further  circuit design. The quantum gates $U_1,\cdots,U_{n-k}$  are designed as 
$U_i=\prod\limits_{j=1}^{n}X^{L_{ji}}_{j}$,
where $X^1_j$ denotes the application of  an $X$ gate  on the $j$-th qubit and  $X^0_j$ denotes the application of  identity  on the $j$-th qubit. Consider a quantum  state $|x_1x_2\cdots x_n\rangle$ in the solution space of the loosened problem,  which  can also be represented as $|S_1S_2\cdots S_{n-k}\rangle_{\rm{\mathbb{S}}}$, we observe that,
\begin{align}
    &U_i|S_1S_2\cdots S_{n-k}\rangle_{\rm{\mathbb{S}}}\notag\\
	=&U_i|x_1x_2\cdots x_n\rangle\notag\\
	=&|(x_1\oplus L_{1i})(x_2\oplus L_{2i})\cdots (x_n\oplus L_{ni})\rangle\notag\\
	=&|S_1S_2\cdots S_{i-1} (S_{i}\oplus1) S_{i+1}\:\cdots S_{n-k}\rangle_{\rm{\mathbb{S}}}.
\end{align}
Consequently, applying $U_i$   flips the $i$-th qubit in the $\mathbb{S}$ representation, effectively implementing a Pauli $X$ gate on that qubit. Subsequently, we define the parameterized quantum gates $G_x(\theta_1,1),\cdots,G_x(\theta_{n-k},n-k)$ as:
\begin{align}
    G_x(\theta_i, i) =&\cos\left(\frac{\theta_i}{2}\right)I - i\sin\left(\frac{\theta_i}{2}\right)U_i\notag\\
 =&\cos\left(\frac{\theta_i}{2}\right)I - i\sin\left(\frac{\theta_i}{2}\right)X_i\;\mathrm{in}\;\mathbb{S}\notag\\
 =& R^i_x(\theta_i)\;\mathrm{in}\;\mathbb{S}.\end{align} 
 The $G_x(\theta, i)$ gates can be effectively implemented using single-qubit rotation gates and two-qubit \textsc{cnot} gates (see Subsection S3 A in the Supplementary Information). 
Based on  these basic gates, we propose the enhanced quantum solvers.
\begin{center}
\subsubsection{VQE-based solver}
\end{center}\par
We focus on the most straightforward ansatz, represented by the following formula:
\begin{equation}
    |\psi(\theta)\rangle = \prod_{i=1}^{n-k}G_x(\theta_i, i) |T_1T_2\cdots T_n\rangle
 = \prod_{i=1}^{n-k}R^i_x(\theta_i)|00\cdots0\rangle_{\rm{\mathbb{S}}}.\label{eq:Requal}
\end{equation}
All quantum states represented by this ansatz remain within the solution space of the loosened problem. We observe that
\begin{equation}
    |\psi(\vec{S}\times\pi)\rangle = \prod_{i=1}^{n-k}(-iX_i)^{S_i}|00\cdots 0\rangle_{\mathbb{S}} = (-i)^{\sum_{i=1}^{n-k}S_i}|\vec{S}\rangle_{\mathbb{S}},
\end{equation}
which is equivalent to the state represented by $\vec{S}$ in Equation~\ref{eq:form}, differing only by a global phase factor. Therefore, each potential solution can be characterized by a set of parameters, enabling us to find solutions through energy minimization. The expectation value of the Hamiltonian for the VQE-based solver can be numerically computed  by multiplying $4\times4$ matrices and vectors, which enables simulation of our VQE-based solver on SAT problems with up to 150 variables (see Subsection S3 D in the Supplementary Information ).\par
The phenomenon of barren plateaus has been extensively studied in quantum optimization, where the cost functions of parameterized quantum circuits exhibit regions in which gradients vanish exponentially with the number of qubits. This characteristic poses significant challenges for optimization algorithms in identifying directions for improvement, thereby reducing the efficiency of quantum algorithms. Recent studies have focused on mitigating and understanding the impact of barren plateaus. The emergence of barren plateaus is influenced by several factors, including complexity of the quantum circuit \cite{cerezo2021cost}, ansatz expressibility \cite{holmes2022connecting},  degree of entanglement in the quantum states \cite{ortiz2021entanglement}, noise \cite{wang2021noise} and structure of the cost function \cite{uvarov2021barren,cerezo2021cost}. \par
Furthermore, research indicates that the presence of barren plateaus is closely related to the 2-design characteristics of circuits \cite{mcclean2018barren} and can be assessed through the gradient variance of the loss function \cite{cerezo2021cost}. We demonstrate that for positive one-in-three SAT problems, our enhanced VQE-based solver is free from barren plateaus if appropriate forms of $L$ and $T$ in Equation~\ref{eq:form} are chosen, as detailed in Subsection~\ref{barren}. This characteristic indicates that our VQE-based solver is trainable, thereby enhancing its feasibility.\par
\subsubsection{QAOA-based solver}
The initial Hamiltonian is chosen as:
\begin{equation}
    H_B=\sum\limits_{i=1}^{n-k}\sigma^i_x\;in\; \mathbb{S}
=\sum\limits_{i=1}^{n-k}U_i.
\end{equation} 
The initial state is correspondingly chosen as:
\begin{align}
    |\psi_0\rangle
    =&\frac{1}{\sqrt{2^{n-k}}}\sum\limits_{S}|S_1S_2\cdots S_{n-k}\rangle_{\mathbb{S}}\notag\\
    =&\frac{1}{\sqrt{2^{n-k}}}\sum\limits_{\mathrm{solution \; space}}|x_1x_2\cdots x_n\rangle.
\end{align}
This state is a superposition of all solutions to the loosened problem and serves as the eigenvector of all $U_i$ with an eigenvalue of 1. Consequently, $|\psi_0\rangle $ is the eigenstate of 
 $H_B$ with the largest eigenvalue,  corresponding to the mixed state in the QAOA framework. The construction of the initial state $|\psi_0\rangle$ is detailed
 in Subsection S3 A in the Supplementary Information. 
 The evolution operators $V_B$ and $V_C$ can be written as: 
\begin{align}
    	V_B(\gamma)&=e^{-i\gamma H_B}\notag\\
        =&\prod\limits_{i=1}^{n-k}e^{-i\gamma U_i}\notag\\
	=&\prod\limits_{i=1}^{n-k} (\cos(\gamma)I-\mathrm{i}\sin(\gamma)U_i)\notag\\
 =&\prod\limits_{i=1}^{n-k}G_x(2\gamma, i), \label{eq:U0}
\end{align}
and 
 \begin{align}
     	V_C(\beta)=&e^{i\beta H_{\mathrm{RSRA}}}\notag\\
        =&\prod\limits_{p=1}^{m}e^{\mathrm{i}\beta W_{p,1}(\frac{1-\sigma_z^{l_{p,1}}}{2})W_{p,2}(\frac{1-\sigma_z^{l_{p,2}}}{2})}.\label{eq:U1}
 \end{align}
Following the framework of QAOA, the entire circuit with $N_{\mathrm{QAOA}}$ layers is designed as:
\begin{equation}\label{eq:QAOAform}
|\psi_{\beta,\gamma}\rangle=V_B(\gamma_{N_{\mathrm{QAOA}}})V_C(\beta_{N_{\mathrm{QAOA}}})\cdots V_B(\gamma_1)V_C(\beta_1)|\psi_0\rangle.
\end{equation}
Classical optimization techniques are employed to determine the values of the  $2N_{\mathrm{QAOA}}$ parameters that minimize $\langle \psi_{\beta,\gamma}|H_{\mathrm{RSRA}}|\psi_{\beta,\gamma}\rangle$. Subsequently, measuring $|\psi_{\beta,\gamma}\rangle$ across all qubits yields a solution to the SAT problem with considerable possibility. \par
The initial state is a mixed state within the solution space of the loosened problem. It has been shown that $G_x$ gates are equivalent to  $R_x$ gates in the $\mathbb{S}$ representation. This equivalence indicates that the $V_B$ gates, which can be decomposed into $G_x$ gates, preserve the quantum state within the solution space. Furthermore,  since the $V_c$ gates solely modify the phases, the entire circuit preserves the solution space of the loosened problem.\par
\subsubsection{QAA-based solver}
To establish a lower bound on the performance of our method, we introduce the QAA-based solver. The QAA-based solver can be viewed as a modified version of the QAOA-based solver that employs empirically selected parameters according to the QAA principles instead of classical optimization. It employs the same Hamiltonian, quantum circuit structure, and initial state as the  QAOA-based solver, ensuring that the quantum state remains within the solution space of the loosened problem.   The ansatz with $N_{\mathrm{QAA}}$ layers can be represented  as:
\begin{equation}\label{eq:QAAform}
|\psi_{\beta,\gamma}\rangle=V_B(\gamma_{N_{\mathrm{QAA}}})V_C(\beta_{N_{\mathrm{QAA}}})\cdots V_B(\gamma_1)V_C(\beta_1)|\psi_0\rangle,
\end{equation}
where $
	\beta_i=cf(\frac{i-1}{N_{\mathrm{QAA}}-1}),\: \gamma_i=c(1-f(\frac{i-1}{N_{\mathrm{QAA}}-1}))$, 
for $i \in  \{1,2, \cdots,N_{\mathrm{QAA}}\}$, where  $c$ is a  constant and $f$ is the scheduling function. According to the quantum adiabatic theorem, the quantum state will evolve toward the eigenstate corresponding to the largest eigenvalue of $-H_{\mathrm{RSRA}}$. Therefore, the solvability of the SAT problem can be determined by assessing whether the final energy approaches zero. The QAA-based solver provides a lower bound on the success probability of the QAOA-based solvers and offers more reliable estimates for scaling behavior. \par
\subsection{Comparison of quantum resources with existing algorithms}
\begin{table*}
\centering
\sffamily 
		\caption{\textbf{Comparison of qubit resources, gate resources and dimensionalities of the search space for different one-in-three SAT solvers in a single run.}  The numbers of required layers for the QAA- and QAOA-based solver are all labeled as $N$. All the enhanced solvers demonstrate advantage in  qubit usage and search space dimensionality, while maintaining moderate quantum gate counts that scales quadratically with the problem size.} 	\label{tabcompare}
    \renewcommand{\arraystretch}{2.6}     
        \setlength{\tabcolsep}{6pt}
	\resizebox{0.7\textwidth}{!}{
    \begin{tabular}{>{\huge}c>{\huge}c>{\huge}c>{\huge}c>{\huge}c>{\huge}c}
    \toprule[2pt]
&\multicolumn{2}{c}{\huge\textbf{Enhanced with the RSRA}}&\hspace{4pt}&\multicolumn{2}{c}{\huge\textbf{Original}}\\
\cmidrule[1pt]{2-3}\cmidrule[1pt]{5-6}
&\textbf{VQE}&\textbf{QAA,QAOA}&&\textbf{VQE}&\textbf{QAA,QAOA}\\
\midrule[1.5pt]
qubit & $|G|\approx 0.5n$&$|G|\approx 0.5n$&&$n$&$n$\\
single-qubit gate&$n-k$&$N(n-k)$&&poly($n$)&$N(n+3m)$\\
two-qubit gate&$2(n-k)(n-1)$&$N(2(n-k)(n-1)+m)$&&poly($n$)&$3Nm$\\
total &$\mathcal{O}(n(n-k))$&$\mathcal{O}(Nn(n-k))$&&poly($n$)&$\mathcal{O}(Nn)$\\
dimensionality&$2^{n-k}=\mathcal{O}(2^{n-m})$&$2^{n-k}=\mathcal{O}(2^{n-m})$&&$2^n$&$2^n$\\
    \bottomrule[2pt]
		\end{tabular}
        }
\end{table*}
We compare the quantum resources required to solve the one-in-three SAT problem with and without classical reduction, focusing on the resources needed for a single execution. The results are summarized in Table~\ref{tabcompare} (see Section S4 in the  Supplementary Information  for detailed calculation). The enhanced quantum solvers exhibit the following key highlights:
\begin{itemize}
    \item The dimensionality of the search space is significantly reduced from $2^n$ to $\mathcal{O}(2^{n-m})$  for all  enhanced solvers.
	\item 	Qubit resources can be reduced from $n$ to $|G|$, which is approximately $0.5n$ in the critical region, across all enhanced solvers.
    \item  Gate resources in the enhanced VQE-based solver  scale quadratically with the problem size, while the gate resources in VQE-based solvers without reduction face challenges in determining the necessary size of the variational space.
    \item Gate resources in each layer of the enhanced QAA- and QAOA-based solver are quadratically related to the problem size.
\end{itemize}
The moderate quantum resource requirements suggest that the enhanced one-in-three SAT solvers have  potential applicability  to real quantum devices in the NISQ era.
\subsection{Absence of barren plateaus in VQE-based solver} \label{barren}
As noted in \cite{cerezo2021cost}, the phenomenon of barren plateaus is closely associated with the variance of the gradient of the loss function in parameterized quantum circuits. The convergence process is free from barren plateaus if the variance does not decay exponentially with increasing problem size.\par 
As demonstrated in Section S2 in the Supplementary Information, each column of the matrix $L$ contains at least one unique element within its corresponding row. 
In a positive one-in-three SAT problem, the initial vector $T$  can be replaced by an all-ones vector, which serves as a solution to the loosened problem. Consequently, the parameterized state can be chosen as follows: \par
\begin{align}\label{eq:positive ansatz}
	|\psi (\theta)\rangle&=\prod\limits_{i=1}^{n-k}R_x^i(\theta_i)|00\cdots0\rangle_\mathbb{S}\notag\\
 &=\prod\limits_{i=1}^{n-k}(\cos(\frac{\theta_i}{2})I-\mathrm{i}\sin(\frac{\theta_i}{2})U_i)|11\cdots 1\rangle.
\end{align} 
In a positive SAT problem, the loss function can be written as:\par
\begin{equation}\label{eq:Hpositive}
	H=\sum\limits_{k=1}^{m} \frac{1}{4}
	(1-\sigma_z^{l_{p,1}})(1-\sigma_z^{l_{p,2}}).
\end{equation}
 Without loss of generality, we consider the $v$-th parameter corresponding to the $v$-th column of $L$.
To demonstrate the absence of barren plateaus in the SAT-based solver, we calculate $\mathrm{var}(\frac{\partial \langle  H\rangle}{\partial \theta_v})$ and prove that it does not approach zero as $n$ increases. Let $\overline{\square }$ denote the average over the parameter space, while $\langle\square \rangle$ represents the expectation of an observable.  \par
The proof consists of two parts.  In the first part, we demonstrate that $\langle H\rangle=A\cos(\theta_v)+C$, where $A$ and $C$ are functions of the other parameters. Consequently, $\frac{\partial  \langle  H\rangle}{\partial \theta_v}=-A\sin(\theta_v)$. Thus, $\overline{\frac{\partial \langle H\rangle}{\partial \theta_v}}=0$ and $\mathrm{var}(\frac{\partial  \langle H\rangle}{\partial \theta_v})=\overline{(\frac{\partial  \langle H\rangle}{\partial \theta_v})^2}=\overline{A^2} \;\overline{\sin(\theta_v)^2}=\overline{\frac{A^2}{2}}\geq \frac{\overline{A}^2}{2}$
, where $\overline{A^2},\overline{A}$  are  the averages of $A^2$ and $A$ under different values of other parameters. In the second part, we prove that $\overline{A}\geq \frac{1}{4}$ if each column of the matrix
$L$ contains at least one element, which is the only nonzero element in its corresponding row. Combining these two parts, we prove that $\mathrm{var}(\frac{\partial \langle  H\rangle}{\partial \theta_v})\geq\frac{1}{32}$,  indicating that the enhanced VQE-based solver is free from barre plateaus when addressing one-in-three SAT problems (see Section S7 in the Supplementary Information  for details).\par
We also conduct numerical simulations to assess the influence of $\frac{m}{n}$ ratio and problem size $n$ on the variance of $\frac{\partial \langle H\rangle}{\partial \theta_v}$. For each ratio $\frac{m}{n}$ and problem size $n$, we randomly generate 1000 one-in-three SAT problems and construct the corresponding ansatz. In each ansätze, we randomly select a parameter as the partial parameter $\theta_v$ and use parameter shift to compute $\frac{\partial \langle H\rangle}{\partial \theta_v}$ for 100 trials, and then calculate their variance. As shown in Fig. S12 in the Supplementary Information, the average variance of $\frac{\partial \langle H\rangle}{\partial \theta_v}$ remains nearly constant across different values of $n$, while it increases rapidly as $\frac{m}{n}$ rises from 0.476 to 0.726, thereby confirming that our solver is free from barren plateaus.

\section*{Acknowledgments}
We thank J. Zhang, JH. Pi for discussions and comments on the manuscript. S. W. acknowledges  Beijing Nova Program under Grants No. 20230484345  and 20240484609; We  acknowledges the National Natural Science Foundation of China under grant No. 62471046.
\section*{Author contributions}
Q. L. and S. W. developed the theoretical framework and executed the
experiments. S. W., J. Z. and G. L. L. supervised the work. Q. L., S. W.,K. L., M. Z, B. Y, P. G., H. Z.and J. Z. contribute to the numerical simulation. All authors contributed in the preparation of the
manuscript. 

\section*{Competing interests}
The authors declare no competing interests.

\section*{Data availability} 
The data that support the findings of this work is available from the authors
on reasonable request.

\section*{Code availability}
The code that supports the findings of this work is available from the authors
on reasonable request.

\bibliographystyle{naturemag}
\citestyle{nature}
\bibliography{Nature.bib}

\begin{thebibliography}{10}
\expandafter\ifx\csname url\endcsname\relax
  \def\url#1{\texttt{#1}}\fi
\expandafter\ifx\csname urlprefix\endcsname\relax\def\urlprefix{URL }\fi
\providecommand{\bibinfo}[2]{#2}
\providecommand{\eprint}[2][]{\url{#2}}

\bibitem{farhi2014quantum}
\bibinfo{author}{Farhi, E.}, \bibinfo{author}{Goldstone, J.} \&
  \bibinfo{author}{Gutmann, S.}
\newblock \bibinfo{title}{A quantum approximate optimization algorithm}.
\newblock \emph{\bibinfo{journal}{arXiv preprint arXiv:1411.4028}}
  (\bibinfo{year}{2014}).
\newblock \urlprefix\url{https://doi.org/10.48550/arXiv.1411.4028}.

\bibitem{peruzzo2014variational}
\bibinfo{author}{Peruzzo, A.} \emph{et~al.}
\newblock \bibinfo{title}{A variational eigenvalue solver on a photonic quantum
  processor}.
\newblock \emph{\bibinfo{journal}{Nat. Commun.}} \textbf{\bibinfo{volume}{5}},
  \bibinfo{pages}{4213} (\bibinfo{year}{2014}).
\newblock \urlprefix\url{https://doi.org/10.1038/ncomms5213}.

\bibitem{Preskill2018}
\bibinfo{author}{Preskill, J.}
\newblock \bibinfo{title}{Quantum {C}omputing in the {NISQ} era and beyond}.
\newblock \emph{\bibinfo{journal}{Quantum}} \textbf{\bibinfo{volume}{2}},
  \bibinfo{pages}{79} (\bibinfo{year}{2018}).
\newblock \urlprefix\url{https://doi.org/10.22331/q-2018-08-06-79}.

\bibitem{Google2019Supremacy}
\bibinfo{author}{Arute, F.} \emph{et~al.}
\newblock \bibinfo{title}{Quantum supremacy using a programmable
  superconducting processor}.
\newblock \emph{\bibinfo{journal}{Nature}} \textbf{\bibinfo{volume}{574}},
  \bibinfo{pages}{505--510} (\bibinfo{year}{2019}).
\newblock \urlprefix\url{https://doi.org/10.1038/s41586-019-1666-5}.

\bibitem{Wu2021Strong}
\bibinfo{author}{Wu, Y.} \emph{et~al.}
\newblock \bibinfo{title}{Strong quantum computational advantage using a
  superconducting quantum processor}.
\newblock \emph{\bibinfo{journal}{Phys. Rev. Lett.}}
  \textbf{\bibinfo{volume}{127}}, \bibinfo{pages}{180501}
  (\bibinfo{year}{2021}).
\newblock
  \urlprefix\url{https://link.aps.org/doi/10.1103/PhysRevLett.127.180501}.

\bibitem{Kim2023Evidence}
\bibinfo{author}{Kim, Y.} \emph{et~al.}
\newblock \bibinfo{title}{Evidence for the utility of quantum computing before
  fault tolerance}.
\newblock \emph{\bibinfo{journal}{Nature}} \textbf{\bibinfo{volume}{618}},
  \bibinfo{pages}{500--505} (\bibinfo{year}{2023}).
\newblock \urlprefix\url{https://doi.org/10.1038/s41586-023-06096-3}.

\bibitem{Acharya2023Suppressing}
\bibinfo{author}{Acharya, R.} \emph{et~al.}
\newblock \bibinfo{title}{Suppressing quantum errors by scaling a surface code
  logical qubit}.
\newblock \emph{\bibinfo{journal}{Nature}} \textbf{\bibinfo{volume}{614}},
  \bibinfo{pages}{676--681} (\bibinfo{year}{2023}).
\newblock \urlprefix\url{https://doi.org/10.1038/s41586-022-05434-1}.

\bibitem{Yu2023Nature}
\bibinfo{author}{Ni, Z.} \emph{et~al.}
\newblock \bibinfo{title}{Beating the break-even point with a
  discrete-variable-encoded logical qubit}.
\newblock \emph{\bibinfo{journal}{Nature}} \textbf{\bibinfo{volume}{616}},
  \bibinfo{pages}{56--60} (\bibinfo{year}{2023}).
\newblock \urlprefix\url{https://doi.org/10.1038/s41586-023-05784-4}.

\bibitem{Sivak2023Nature}
\bibinfo{author}{Sivak, V.~V.} \emph{et~al.}
\newblock \bibinfo{title}{Real-time quantum error correction beyond
  break-even}.
\newblock \emph{\bibinfo{journal}{Nature}} \textbf{\bibinfo{volume}{616}},
  \bibinfo{pages}{50--55} (\bibinfo{year}{2023}).
\newblock \urlprefix\url{https://doi.org/10.1038/s41586-023-05782-6}.

\bibitem{Lukin2024Nature}
\bibinfo{author}{Bluvstein, D.} \emph{et~al.}
\newblock \bibinfo{title}{Logical quantum processor based on reconfigurable
  atom arrays}.
\newblock \emph{\bibinfo{journal}{Nature}} \textbf{\bibinfo{volume}{626}},
  \bibinfo{pages}{58--65} (\bibinfo{year}{2024}).
\newblock \urlprefix\url{https://doi.org/10.1038/s41586-023-06927-3}.

\bibitem{boulebnane2024solving}
\bibinfo{author}{Boulebnane, S.} \& \bibinfo{author}{Montanaro, A.}
\newblock \bibinfo{title}{Solving boolean satisfiability problems with the
  quantum approximate optimization algorithm}.
\newblock \emph{\bibinfo{journal}{PRX Quantum}} \textbf{\bibinfo{volume}{5}},
  \bibinfo{pages}{030348} (\bibinfo{year}{2024}).
\newblock \urlprefix\url{https://doi.org/10.1103/PRXQuantum.5.030348}.

\bibitem{shaydulin2024evidence}
\bibinfo{author}{Shaydulin, R.} \emph{et~al.}
\newblock \bibinfo{title}{Evidence of scaling advantage for the quantum
  approximate optimization algorithm on a classically intractable problem}.
\newblock \emph{\bibinfo{journal}{Sci. Adv.}} \textbf{\bibinfo{volume}{10}},
  \bibinfo{pages}{eadm6761} (\bibinfo{year}{2024}).
\newblock \urlprefix\url{https://www.science.org/doi/10.1126/sciadv.adm6761}.

\bibitem{biere1999symbolic}
\bibinfo{author}{Biere, A.}, \bibinfo{author}{Cimatti, A.},
  \bibinfo{author}{Clarke, E.~M.}, \bibinfo{author}{Fujita, M.} \&
  \bibinfo{author}{Zhu, Y.}
\newblock \bibinfo{title}{Symbolic model checking using sat procedures instead
  of bdds}.
\newblock In \emph{\bibinfo{booktitle}{Proceedings of the 36th annual ACMIEEE
  Design Automation Conference}}, \bibinfo{pages}{317--320}
  (\bibinfo{year}{1999}).
\newblock \urlprefix\url{https://ieeexplore.ieee.org/document/781333}.

\bibitem{bradley2011sat}
\bibinfo{author}{Bradley, A.~R.}
\newblock \bibinfo{title}{Sat-based model checking without unrolling}.
\newblock In \emph{\bibinfo{booktitle}{International Workshop on Verification,
  Model Checking, and Abstract Interpretation}}, \bibinfo{pages}{70--87}
  (\bibinfo{organization}{Springer}, \bibinfo{year}{2011}).
\newblock
  \urlprefix\url{https://link.springer.com/chapter/10.1007/978-3-642-18275-4_7}.

\bibitem{mcmillan2003interpolation}
\bibinfo{author}{McMillan, K.~L.}
\newblock \bibinfo{title}{Interpolation and sat-based model checking}.
\newblock In \emph{\bibinfo{booktitle}{Computer Aided Verification: 15th
  International Conference, CAV 2003, Boulder, CO, USA, July 8-12, 2003.
  Proceedings 15}}, \bibinfo{pages}{1--13} (\bibinfo{organization}{Springer},
  \bibinfo{year}{2003}).
\newblock \urlprefix\url{https://doi.org/10.1007/978-3-540-45069-6\_1}.

\bibitem{amla2005analysis}
\bibinfo{author}{Amla, N.}, \bibinfo{author}{Du, X.},
  \bibinfo{author}{Kuehlmann, A.}, \bibinfo{author}{Kurshan, R.~P.} \&
  \bibinfo{author}{McMillan, K.~L.}
\newblock \bibinfo{title}{An analysis of sat-based model checking techniques in
  an industrial environment}.
\newblock In \emph{\bibinfo{booktitle}{Correct Hardware Design and Verification
  Methods: 13th IFIP WG 10.5 Advanced Research Working Conference, CHARME 2005,
  Saarbr{\"u}cken, Germany, October 3-6, 2005. Proceedings 13}},
  \bibinfo{pages}{254--268} (\bibinfo{organization}{Springer},
  \bibinfo{year}{2005}).
\newblock \urlprefix\url{https://doi.org/10.1007/11560548\_20}.

\bibitem{zulkoski2017combining}
\bibinfo{author}{Zulkoski, E.} \emph{et~al.}
\newblock \bibinfo{title}{Combining sat solvers with computer algebra systems
  to verify combinatorial conjectures}.
\newblock \emph{\bibinfo{journal}{J. Autom. Reason.}}
  \textbf{\bibinfo{volume}{58}}, \bibinfo{pages}{313--339}
  (\bibinfo{year}{2017}).
\newblock \urlprefix\url{https://doi.org/10.1007/s10817-016-9396-y}.

\bibitem{bejar2007regular}
\bibinfo{author}{B{\'e}jar, R.}, \bibinfo{author}{Manya, F.},
  \bibinfo{author}{Cabiscol, A.}, \bibinfo{author}{Fern{\'a}ndez, C.} \&
  \bibinfo{author}{Gomes, C.}
\newblock \bibinfo{title}{Regular-sat: A many-valued approach to solving
  combinatorial problems}.
\newblock \emph{\bibinfo{journal}{Discret. Appl. Math.}}
  \textbf{\bibinfo{volume}{155}}, \bibinfo{pages}{1613--1626}
  (\bibinfo{year}{2007}).
\newblock
  \urlprefix\url{https://www.sciencedirect.com/science/article/pii/S0166218X06004653}.

\bibitem{banbara2010generating}
\bibinfo{author}{Banbara, M.}, \bibinfo{author}{Matsunaka, H.},
  \bibinfo{author}{Tamura, N.} \& \bibinfo{author}{Inoue, K.}
\newblock \bibinfo{title}{Generating combinatorial test cases by efficient sat
  encodings suitable for cdcl sat solvers}.
\newblock In \emph{\bibinfo{booktitle}{Logic for Programming, Artificial
  Intelligence, and Reasoning: 17th International Conference, LPAR-17,
  Yogyakarta, Indonesia, October 10-15, 2010. Proceedings 17}},
  \bibinfo{pages}{112--126} (\bibinfo{organization}{Springer},
  \bibinfo{year}{2010}).
\newblock \urlprefix\url{https://doi.org/10.1007/978-3-642-16242-8\_9}.

\bibitem{han2010making}
\bibinfo{author}{Han, H.}, \bibinfo{author}{Somenzi, F.} \&
  \bibinfo{author}{Jin, H.}
\newblock \bibinfo{title}{Making deduction more effective in sat solvers}.
\newblock \emph{\bibinfo{journal}{IEEE Transactions on Comput. Des. Integr.
  Circuits Syst.}} \textbf{\bibinfo{volume}{29}}, \bibinfo{pages}{1271--1284}
  (\bibinfo{year}{2010}).
\newblock \urlprefix\url{https://ieeexplore.ieee.org/document/5512695}.

\bibitem{armando2005sat}
\bibinfo{author}{Armando, A.}, \bibinfo{author}{Castellini, C.},
  \bibinfo{author}{Giunchiglia, E.}, \bibinfo{author}{Giunchiglia, F.} \&
  \bibinfo{author}{Tacchella, A.}
\newblock \emph{\bibinfo{title}{SAT-Based Decision Procedures for Automated
  Reasoning: A Unifying Perspective}}, \bibinfo{pages}{46--58}
  (\bibinfo{publisher}{Springer Berlin Heidelberg}, \bibinfo{address}{Berlin,
  Heidelberg}, \bibinfo{year}{2005}).
\newblock \urlprefix\url{https://doi.org/10.1007/978-3-540-32254-2\_4}.

\bibitem{soos2009extending}
\bibinfo{author}{Soos, M.}, \bibinfo{author}{Nohl, K.} \&
  \bibinfo{author}{Castelluccia, C.}
\newblock \bibinfo{title}{Extending sat solvers to cryptographic problems}.
\newblock In \emph{\bibinfo{booktitle}{International Conference on Theory and
  Applications of Satisfiability Testing}}, \bibinfo{pages}{244--257}
  (\bibinfo{organization}{Springer}, \bibinfo{year}{2009}).
\newblock \urlprefix\url{https://doi.org/10.1007/978-3-642-02777-2\_24}.

\bibitem{otpuschennikov2016encoding}
\bibinfo{author}{Otpuschennikov, I.}, \bibinfo{author}{Semenov, A.},
  \bibinfo{author}{Gribanova, I.}, \bibinfo{author}{Zaikin, O.} \&
  \bibinfo{author}{Kochemazov, S.}
\newblock \bibinfo{title}{Encoding cryptographic functions to sat using
  transalg system}.
\newblock In \emph{\bibinfo{booktitle}{Proceedings of the Twenty-Second
  European Conference on Artificial Intelligence}}, ECAI'16,
  \bibinfo{pages}{1594–1595} (\bibinfo{publisher}{IOS Press},
  \bibinfo{address}{NLD}, \bibinfo{year}{2016}).
\newblock \urlprefix\url{https://doi.org/10.3233/978-1-61499-672-9-1594}.

\bibitem{massacci2000logical}
\bibinfo{author}{Massacci, F.} \& \bibinfo{author}{Marraro, L.}
\newblock \bibinfo{title}{Logical cryptanalysis as a sat problem}.
\newblock \emph{\bibinfo{journal}{J. Autom. Reason.}}
  \textbf{\bibinfo{volume}{24}}, \bibinfo{pages}{165--203}
  (\bibinfo{year}{2000}).
\newblock \urlprefix\url{https://doi.org/10.1023/A:1006326723002}.

\bibitem{mironov2006applications}
\bibinfo{author}{Mironov, I.} \& \bibinfo{author}{Zhang, L.}
\newblock \bibinfo{title}{Applications of sat solvers to cryptanalysis of hash
  functions}.
\newblock In \emph{\bibinfo{booktitle}{Theory and Applications of
  Satisfiability Testing-SAT 2006: 9th International Conference, Seattle, WA,
  USA, August 12-15, 2006. Proceedings 9}}, \bibinfo{pages}{102--115}
  (\bibinfo{organization}{Springer}, \bibinfo{year}{2006}).
\newblock \urlprefix\url{https://eprint.iacr.org/2006/254}.

\bibitem{cook1971}
\bibinfo{author}{Cook, S.~A.}
\newblock \bibinfo{title}{The complexity of theorem-proving procedures}.
\newblock In \emph{\bibinfo{booktitle}{Proceedings of the Third Annual ACM
  Symposium on Theory of Computing}}, STOC '71, \bibinfo{pages}{151–158}
  (\bibinfo{publisher}{Association for Computing Machinery},
  \bibinfo{address}{New York, NY, USA}, \bibinfo{year}{1971}).
\newblock \urlprefix\url{https://doi.org/10.1145/800157.805047}.

\bibitem{nusslein2023solving}
\bibinfo{author}{N{\"u}{\ss}lein, J.}, \bibinfo{author}{Zielinski, S.},
  \bibinfo{author}{Gabor, T.}, \bibinfo{author}{Linnhoff-Popien, C.} \&
  \bibinfo{author}{Feld, S.}
\newblock \bibinfo{title}{Solving (max) 3-sat via quadratic unconstrained
  binary optimization}.
\newblock In \emph{\bibinfo{booktitle}{International Conference on
  Computational Science}}, \bibinfo{pages}{34--47}
  (\bibinfo{organization}{Springer}, \bibinfo{year}{2023}).
\newblock \urlprefix\url{https://doi.org/10.1007/978-3-031-36030-5\_3}.

\bibitem{albash2018adiabatic}
\bibinfo{author}{Albash, T.} \& \bibinfo{author}{Lidar, D.~A.}
\newblock \bibinfo{title}{Adiabatic quantum computation}.
\newblock \emph{\bibinfo{journal}{Rev. Mod. Phys.}}
  \textbf{\bibinfo{volume}{90}}, \bibinfo{pages}{015002}
  (\bibinfo{year}{2018}).
\newblock
  \urlprefix\url{https://link.aps.org/doi/10.1103/RevModPhys.90.015002}.

\bibitem{hogg2003adiabatic}
\bibinfo{author}{Hogg, T.}
\newblock \bibinfo{title}{Adiabatic quantum computing for random satisfiability
  problems}.
\newblock \emph{\bibinfo{journal}{Phys. Rev. A}} \textbf{\bibinfo{volume}{67}},
  \bibinfo{pages}{022314} (\bibinfo{year}{2003}).
\newblock \urlprefix\url{https://link.aps.org/doi/10.1103/PhysRevA.67.022314}.

\bibitem{wang2016ultrafast}
\bibinfo{author}{Wang, H.} \& \bibinfo{author}{Wu, L.-A.}
\newblock \bibinfo{title}{Ultrafast adiabatic quantum algorithm for the
  np-complete exact cover problem}.
\newblock \emph{\bibinfo{journal}{Sci. Rep.}} \textbf{\bibinfo{volume}{6}},
  \bibinfo{pages}{22307} (\bibinfo{year}{2016}).
\newblock \urlprefix\url{https://doi.org/10.1038/srep22307}.

\bibitem{farhi2001quantum}
\bibinfo{author}{Farhi, E.} \emph{et~al.}
\newblock \bibinfo{title}{A quantum adiabatic evolution algorithm applied to
  random instances of an np-complete problem}.
\newblock \emph{\bibinfo{journal}{Science}} \textbf{\bibinfo{volume}{292}},
  \bibinfo{pages}{472--475} (\bibinfo{year}{2001}).
\newblock \urlprefix\url{https://www.science.org/doi/10.1126/science.1057726}.

\bibitem{young2008size}
\bibinfo{author}{Young, A.~P.}, \bibinfo{author}{Knysh, S.} \&
  \bibinfo{author}{Smelyanskiy, V.~N.}
\newblock \bibinfo{title}{Size dependence of the minimum excitation gap in the
  quantum adiabatic algorithm}.
\newblock \emph{\bibinfo{journal}{Phys. Rev. Lett.}}
  \textbf{\bibinfo{volume}{101}}, \bibinfo{pages}{170503}
  (\bibinfo{year}{2008}).
\newblock
  \urlprefix\url{https://link.aps.org/doi/10.1103/PhysRevLett.101.170503}.

\bibitem{young2010first}
\bibinfo{author}{Young, A.}, \bibinfo{author}{Knysh, S.} \&
  \bibinfo{author}{Smelyanskiy, V.}
\newblock \bibinfo{title}{First-order phase transition in the quantum adiabatic
  algorithm}.
\newblock \emph{\bibinfo{journal}{Phys. Rev. Lett.}}
  \textbf{\bibinfo{volume}{104}}, \bibinfo{pages}{020502}
  (\bibinfo{year}{2010}).
\newblock
  \urlprefix\url{https://link.aps.org/doi/10.1103/PhysRevLett.104.020502}.

\bibitem{golden2023quantum}
\bibinfo{author}{Golden, J.}, \bibinfo{author}{B{\"a}rtschi, A.},
  \bibinfo{author}{O'Malley, D.} \& \bibinfo{author}{Eidenbenz, S.}
\newblock \bibinfo{title}{The quantum alternating operator ansatz for
  satisfiability problems}.
\newblock In \emph{\bibinfo{booktitle}{2023 IEEE International Conference on
  Quantum Computing and Engineering (QCE)}}, vol.~\bibinfo{volume}{1},
  \bibinfo{pages}{307--312} (\bibinfo{organization}{IEEE},
  \bibinfo{year}{2023}).
\newblock
  \urlprefix\url{https://doi.ieeecomputersociety.org/10.1109/QCE57702.2023.00042}.

\bibitem{mandl2024amplitude}
\bibinfo{author}{Mandl, A.}, \bibinfo{author}{Barzen, J.},
  \bibinfo{author}{Bechtold, M.}, \bibinfo{author}{Leymann, F.} \&
  \bibinfo{author}{Wild, K.}
\newblock \bibinfo{title}{Amplitude amplification-inspired qaoa: Improving the
  success probability for solving 3sat}.
\newblock \emph{\bibinfo{journal}{QuantumSci. Technol.}}
  \textbf{\bibinfo{volume}{9}}, \bibinfo{pages}{015028} (\bibinfo{year}{2024}).
\newblock \urlprefix\url{https://dx.doi.org/10.1088/2058-9565/ad141d}.

\bibitem{bengtsson2020improved}
\bibinfo{author}{Bengtsson, A.} \emph{et~al.}
\newblock \bibinfo{title}{Improved success probability with greater circuit
  depth for the quantum approximate optimization algorithm}.
\newblock \emph{\bibinfo{journal}{Phys. Rev. Appl.}}
  \textbf{\bibinfo{volume}{14}}, \bibinfo{pages}{034010}
  (\bibinfo{year}{2020}).
\newblock
  \urlprefix\url{https://link.aps.org/doi/10.1103/PhysRevApplied.14.034010}.

\bibitem{raymond2007phase}
\bibinfo{author}{Raymond, J.}, \bibinfo{author}{Sportiello, A.} \&
  \bibinfo{author}{Zdeborov{\'a}, L.}
\newblock \bibinfo{title}{Phase diagram of the 1-in-3 satisfiability problem}.
\newblock \emph{\bibinfo{journal}{Phys. Rev. E}} \textbf{\bibinfo{volume}{76}},
  \bibinfo{pages}{011101} (\bibinfo{year}{2007}).
\newblock \urlprefix\url{https://link.aps.org/doi/10.1103/PhysRevE.76.011101}.

\bibitem{sorensson2005minisat}
\bibinfo{author}{Sorensson, N.} \& \bibinfo{author}{Een, N.}
\newblock \bibinfo{title}{Minisat v1. 13-a sat solver with conflict-clause
  minimization}.
\newblock \emph{\bibinfo{journal}{SAT}} \textbf{\bibinfo{volume}{2005}},
  \bibinfo{pages}{1--2} (\bibinfo{year}{2005}).
\newblock \urlprefix\url{http://minisat.se/downloads/MiniSat_v1.13_short.pdf}.

\bibitem{sorensson2010minisat}
\bibinfo{author}{S{\"o}rensson, N.}
\newblock \bibinfo{title}{Minisat 2.2 and minisat++ 1.1}.
\newblock \emph{\bibinfo{journal}{A short description in SAT Race}}
  \textbf{\bibinfo{volume}{2010}} (\bibinfo{year}{2010}).
\newblock \urlprefix\url{https://api.semanticscholar.org/CorpusID:30567720}.

\bibitem{audemard:hal-03299473}
\bibinfo{author}{Audemard, G.} \& \bibinfo{author}{Simon, L.}
\newblock \bibinfo{title}{On the glucose sat solver}.
\newblock \emph{\bibinfo{journal}{Int. J. Artif. Intell. Tools}}
  \textbf{\bibinfo{volume}{27}}, \bibinfo{pages}{1--25} (\bibinfo{year}{2018}).
\newblock \urlprefix\url{https://univ-artois.hal.science/hal-03299473}.

\bibitem{Biere-SAT-Competition-2017-solvers}
\bibinfo{author}{Biere, A.}
\newblock \bibinfo{title}{{CaDiCaL, Lingeling, Plingeling, Treengeling, YalSAT
  Entering the SAT Competition 2017}}.
\newblock In \bibinfo{editor}{Balyo, T.}, \bibinfo{editor}{Heule, M.} \&
  \bibinfo{editor}{J{\"a}rvisalo, M.} (eds.) \emph{\bibinfo{booktitle}{Proc.~of
  {SAT Competition} 2017 -- Solver and Benchmark Descriptions}}, vol.
  \bibinfo{volume}{B-2017-1} of \emph{\bibinfo{series}{Department of Computer
  Science Series of Publications B}}, \bibinfo{pages}{14--15}
  (\bibinfo{publisher}{University of Helsinki}, \bibinfo{year}{2017}).
\newblock
  \urlprefix\url{https://fmv.jku.at/papers/Biere-SAT-Race-2019-solvers.pdf}.

\bibitem{Biere-SAT-Competition-2016-solvers}
\bibinfo{author}{Biere, A.}
\newblock \bibinfo{title}{{Splatz, Lingeling, Plingeling, Treengeling, YalSAT
  Entering the SAT Competition 2016}}.
\newblock In \bibinfo{editor}{Balyo, T.}, \bibinfo{editor}{Heule, M.} \&
  \bibinfo{editor}{J{\"a}rvisalo, M.} (eds.) \emph{\bibinfo{booktitle}{Proc.~of
  {SAT Competition} 2016 -- Solver and Benchmark Descriptions}}, vol.
  \bibinfo{volume}{B-2016-1} of \emph{\bibinfo{series}{Department of Computer
  Science Series of Publications B}}, \bibinfo{pages}{44--45}
  (\bibinfo{publisher}{University of Helsinki}, \bibinfo{year}{2016}).
\newblock \urlprefix\url{https://api.semanticscholar.org/CorpusID:2495003}.

\bibitem{queue2019cadical}
\bibinfo{author}{Biere, A.}
\newblock \bibinfo{title}{Cadical at the sat race 2019}.
\newblock In \emph{\bibinfo{booktitle}{Proceedings of SAT Race 2019 – Solver
  and Benchmark Descriptions}}, vol. \bibinfo{volume}{B-2019-1} of
  \emph{\bibinfo{series}{Department of Computer Science Series of Publications
  B}}, \bibinfo{pages}{8--9} (\bibinfo{publisher}{University of Helsinki},
  \bibinfo{year}{2019}).
\newblock \urlprefix\url{https://api.semanticscholar.org/CorpusID:208155834}.

\bibitem{knuth2000analogue}
\bibinfo{author}{Knuth, D.~E.}
\newblock \bibinfo{title}{Dancing links}.
\newblock \emph{\bibinfo{journal}{arXiv preprint cs/0011047}}
  (\bibinfo{year}{2000}).
\newblock \urlprefix\url{https://arxiv.org/abs/cs/0011047}.

\bibitem{li2017hybrid}
\bibinfo{author}{Li, J.}, \bibinfo{author}{Yang, X.}, \bibinfo{author}{Peng,
  X.} \& \bibinfo{author}{Sun, C.-P.}
\newblock \bibinfo{title}{Hybrid quantum-classical approach to quantum optimal
  control}.
\newblock \emph{\bibinfo{journal}{Phys. Rev. Lett.}}
  \textbf{\bibinfo{volume}{118}}, \bibinfo{pages}{150503}
  (\bibinfo{year}{2017}).
\newblock
  \urlprefix\url{https://link.aps.org/doi/10.1103/PhysRevLett.118.150503}.

\bibitem{mitarai2018quantum}
\bibinfo{author}{Mitarai, K.}, \bibinfo{author}{Negoro, M.},
  \bibinfo{author}{Kitagawa, M.} \& \bibinfo{author}{Fujii, K.}
\newblock \bibinfo{title}{Quantum circuit learning}.
\newblock \emph{\bibinfo{journal}{Phys. Rev. A}} \textbf{\bibinfo{volume}{98}},
  \bibinfo{pages}{032309} (\bibinfo{year}{2018}).
\newblock \urlprefix\url{https://link.aps.org/doi/10.1103/PhysRevA.98.032309}.

\bibitem{brassard2000quantum}
\bibinfo{author}{Brassard, G.}, \bibinfo{author}{Hoyer, P.},
  \bibinfo{author}{Mosca, M.} \& \bibinfo{author}{Tapp, A.}
\newblock \bibinfo{title}{Quantum amplitude amplification and estimation}.
\newblock \emph{\bibinfo{journal}{arXiv preprint quant-ph/0005055}}
  (\bibinfo{year}{2000}).
\newblock \urlprefix\url{https://arxiv.org/abs/quant-ph/0005055}.

\bibitem{schoning1999probabilistic}
\bibinfo{author}{Schoning, T.}
\newblock \bibinfo{title}{A probabilistic algorithm for k-sat and constraint
  satisfaction problems}.
\newblock In \emph{\bibinfo{booktitle}{40th Annual Symposium on Foundations of
  Computer Science (Cat. No. 99CB37039)}}, \bibinfo{pages}{410--414}
  (\bibinfo{organization}{IEEE}, \bibinfo{year}{1999}).
\newblock \urlprefix\url{https://ieeexplore.ieee.org/document/814612}.

\bibitem{bravyi2006efficient}
\bibinfo{author}{Bravyi, S.}
\newblock \bibinfo{title}{Efficient algorithm for a quantum analogue of 2-sat}.
\newblock \emph{\bibinfo{journal}{arXiv preprint quant-ph/0602108}}
  (\bibinfo{year}{2006}).
\newblock \urlprefix\url{https://arxiv.org/abs/quant-ph/0602108}.

\bibitem{cerezo2021cost}
\bibinfo{author}{Cerezo, M.}, \bibinfo{author}{Sone, A.},
  \bibinfo{author}{Volkoff, T.}, \bibinfo{author}{Cincio, L.} \&
  \bibinfo{author}{Coles, P.~J.}
\newblock \bibinfo{title}{Cost function dependent barren plateaus in shallow
  parametrized quantum circuits}.
\newblock \emph{\bibinfo{journal}{Nat. Commun.}} \textbf{\bibinfo{volume}{12}},
  \bibinfo{pages}{1791} (\bibinfo{year}{2021}).
\newblock \urlprefix\url{https://doi.org/10.1038/s41467-021-21728-w}.

\bibitem{holmes2022connecting}
\bibinfo{author}{Holmes, Z.}, \bibinfo{author}{Sharma, K.},
  \bibinfo{author}{Cerezo, M.} \& \bibinfo{author}{Coles, P.~J.}
\newblock \bibinfo{title}{Connecting ansatz expressibility to gradient
  magnitudes and barren plateaus}.
\newblock \emph{\bibinfo{journal}{PRX Quantum}} \textbf{\bibinfo{volume}{3}},
  \bibinfo{pages}{010313} (\bibinfo{year}{2022}).
\newblock \urlprefix\url{https://link.aps.org/doi/10.1103/PRXQuantum.3.010313}.

\bibitem{ortiz2021entanglement}
\bibinfo{author}{Ortiz~Marrero, C.}, \bibinfo{author}{Kieferov{\'a}, M.} \&
  \bibinfo{author}{Wiebe, N.}
\newblock \bibinfo{title}{Entanglement-induced barren plateaus}.
\newblock \emph{\bibinfo{journal}{PRX Quantum}} \textbf{\bibinfo{volume}{2}},
  \bibinfo{pages}{040316} (\bibinfo{year}{2021}).
\newblock \urlprefix\url{https://link.aps.org/doi/10.1103/PRXQuantum.2.040316}.

\bibitem{wang2021noise}
\bibinfo{author}{Wang, S.} \emph{et~al.}
\newblock \bibinfo{title}{Noise-induced barren plateaus in variational quantum
  algorithms}.
\newblock \emph{\bibinfo{journal}{Nat. Commun.}} \textbf{\bibinfo{volume}{12}},
  \bibinfo{pages}{6961} (\bibinfo{year}{2021}).
\newblock \urlprefix\url{https://doi.org/10.1038/s41467-021-27045-6}.

\bibitem{uvarov2021barren}
\bibinfo{author}{Uvarov, A.} \& \bibinfo{author}{Biamonte, J.~D.}
\newblock \bibinfo{title}{On barren plateaus and cost function locality in
  variational quantum algorithms}.
\newblock \emph{\bibinfo{journal}{J. Phys. A:Math. Theor.}}
  \textbf{\bibinfo{volume}{54}}, \bibinfo{pages}{245301}
  (\bibinfo{year}{2021}).
\newblock
  \urlprefix\url{https://iopscience.iop.org/article/10.1088/1751-8121/abfac7}.

\bibitem{mcclean2018barren}
\bibinfo{author}{McClean, J.~R.}, \bibinfo{author}{Boixo, S.},
  \bibinfo{author}{Smelyanskiy, V.~N.}, \bibinfo{author}{Babbush, R.} \&
  \bibinfo{author}{Neven, H.}
\newblock \bibinfo{title}{Barren plateaus in quantum neural network training
  landscapes}.
\newblock \emph{\bibinfo{journal}{Nat. Commun.}} \textbf{\bibinfo{volume}{9}},
  \bibinfo{pages}{4812} (\bibinfo{year}{2018}).
\newblock \urlprefix\url{https://doi.org/10.1038/s41467-018-07090-4}.

\end{thebibliography}

\end{document}